\documentclass{aa}  
\usepackage{graphicx}
\usepackage{txfonts}
\usepackage{textcomp}
\usepackage{longtable}
\usepackage{pdflscape}
\usepackage{natbib}
\usepackage{amsmath}
\usepackage{mathabx}
\usepackage[pdfstartview=FitH,      
            pdfborder={0 0 0},
            colorlinks=true,
            linkcolor=blue,
            citecolor=blue,
            breaklinks=true,       
            bookmarksopen=true,     
            bookmarksnumbered=true  
            ]{hyperref}

\begin{document}
\title{Kepler Object of Interest Network}
\subtitle{II. Photodynamical modelling of Kepler-9 over 8 years of transit observations}
\author{J.~Freudenthal\inst{\ref{inst1}}\thanks{E-mail: jfreude@astro.physik.uni-goettingen.de} \and 
C.~von~Essen\inst{\ref{inst2},\ref{inst1}} \and
S.~Dreizler\inst{\ref{inst1}} \and 
S.~Wedemeyer\inst{\ref{Sven1},\ref{Sven2}} \and
E.~Agol\inst{\ref{Eric1},\ref{Eric2},\ref{Eric3},\ref{Eric4}} \and
B.~M.~Morris\inst{\ref{Eric1}} \and
A.~C.~Becker\inst{\ref{Eric1}} \and
M.~Mallonn\inst{\ref{Matthi1}} \and
S.~Hoyer\inst{\ref{Sergio1},\ref{Sergio2},\ref{Sergio3}} \and
A.~Ofir\inst{\ref{inst1},\ref{Aviv1}} \and
L.~Tal-Or\inst{\ref{inst1},\ref{Lev1}} \and
H.~J.~Deeg\inst{\ref{Sergio1},\ref{Sergio2}} \and
E.~Herrero\inst{\ref{Kike1},\ref{Kike2}} \and
I.~Ribas\inst{\ref{Kike1},\ref{Kike2}} \and
S.~Khalafinejad\inst{\ref{Sara1},\ref{Sara2}} \and
J.~Hern\'andez\inst{\ref{Jesus1}} \and
M.~M.~Rodr\'iguez~S.\inst{\ref{Jesus1}}
}
\institute{Institut f\"{u}r Astrophysik, Georg-August-Universit\"{a}t G\"{o}ttingen, Friedrich-Hund-Platz\,1, 37077 G\"{o}ttingen, Germany\label{inst1} \and 
Stellar Astrophysics Centre, Aarhus University, Ny Munkegade 120, 8000 Aarhus, Denmark\label{inst2} \and
Rosseland Centre for Solar Physics, University of Oslo, P.O. Box 1029 Blindern, N-0315 Oslo, Norway\label{Sven1} \and
Institute of Theoretical Astrophysics, University of Oslo, P.O. Box 1029 Blindern, N-0315 Oslo, Norway\label{Sven2} \and
Astronomy Department, University of Washington, Seattle, WA 98195, United States of America\label{Eric1} \and
Institut d'Astrophysique de Paris, 98 bis Boulevard Arago, Paris 75014, France\label{Eric2} \and
Guggenheim Fellow\label{Eric3} \and
Virtual Planetary Laboratory\label{Eric4} \and
Leibniz-Institut f\"{u}r Astrophysik Potsdam, An der Sternwarte 16, D-14482 Potsdam, Germany\label{Matthi1} \and
Instituto de Astrof\'\i sica de Canarias, C. V\'\i a L\'actea S/N, E-38205 La Laguna, Tenerife, Spain\label{Sergio1} \and
Universidad de La Laguna, Dept. de Astrof\'\i sica, E-38206 La Laguna, Tenerife, Spain\label{Sergio2} \and
Aix Marseille Univ, CNRS, LAM, Laboratoire d'Astrophysique de Marseille, Marseille, France\label{Sergio3} \and
Department of Earth and Planetary Sciences, Weizmann Institute of Science, Rehovot, 76100, Israel\label{Aviv1} \and
School of Geosciences, Raymond and Beverly Sackler Faculty of Exact Sciences, Tel Aviv University, Tel Aviv, 6997801, Israel\label{Lev1} \and
Institut de Ciències de l'Espai (IEEC-CSIC), C/Can Magrans, s/n, Campus UAB, 08193 Bellaterra, Spain\label{Kike1} \and
Institut d’Estudis Espacials de Catalunya (IEEC), C/Gran Capità 2-4, Edif. Nexus, 08034 Barcelona, Spain\label{Kike2} \and
Hamburg Observatory, Hamburg University, Gojenbergsweg 112, 21029 Hamburg, Germany\label{Sara1} \and
Max Planck Institute for Astronomy, K\"onigstuhl 17, 69117 Heidelberg, Germany\label{Sara2} \and
Instituto de Astronom\'ia, UNAM, Campus Ensenada, Carretera Tijuana-Ensenada km 103, 22860 Ensenada, B.C. M\'exico\label{Jesus1}
}
\date{Received 16/05/2018 / Accepted 28/06/2018}

\abstract {The {\it Kepler} Object of Interest Network (KOINet) is a
  multi-site network of telescopes around the globe organised to
  follow up transiting planet candidate KOIs with large transit timing
  variations (TTVs). Its main goal is to complete their TTV curves, as
  the {\it Kepler} telescope no longer observes the original {\it
    Kepler} field.}  {Combining {\it Kepler} and new ground-based
  transit data we improve the modelling of these systems. To this end,
  we have developed a photodynamical model, and we demonstrate its
  performance using the Kepler-9 system as an example.}  {Our
  comprehensive analysis combines the numerical integration of the
  system's dynamics over the time span of the observations along with
  the transit light curve model. This provides a coherent description
  of all observations simultaneously. This model is coupled with a
  Markov chain Monte Carlo algorithm, allowing the exploration of the
  model parameter space.}  {Applied to the Kepler-9 long cadence data,
  short cadence data and 13 new transit observations collected by
  KOINet between the years 2014 to 2017, our modelling provides well
  constrained predictions for the next transits and the system's
  parameters. We have determined the densities of the planets
  Kepler-9b and 9c to the very precise values of
  $\rho_b=0.439\pm0.023\,{\rm g~cm}^{-3}$ and
  $\rho_c=0.322\pm0.017\,{\rm g~cm}^{-3}$.  Our analysis reveals that
  Kepler-9c will stop transiting in about 30 years. This results from
  strong dynamical interactions between Kepler-9b and 9c, near 2:1
  resonance, that leads to a periodic change in inclination.}  {Over
  the next 30 years the inclination of Kepler-9c (-9b) will decrease
  (increase) slowly. This should be measurable by a substantial
  decrease (increase) in the transit duration, in as soon as a few
  years' time. Observations that contradict this prediction might
  indicate the presence of additional objects in this system.  If this
  prediction proves true, this behaviour opens up a unique chance to
  scan the different latitudes of a star: high latitudes with planet c
  and low latitudes with planet b.}   \maketitle

\begin{abstract}

\end{abstract}

\section{Introduction}
One of the outstanding results of the {\it Kepler} mission
\citep{2010Sci...327..977B} is the large number of transiting
multi-planet systems. Prior to  {\it Kepler's} launch it was
shown that the analysis of the dynamical interaction in multi-planet
systems would be feasible offering an independent mass determination
\citep{Holman2005,Agol2005}. This was impressively
confirmed from the first multi-transiting systems
\citep{Holman2010,2011Natur.470...53L} using transit timing
variations (TTVs), that is, deviations from strict periodicity in
planetary transits, caused by non-Keplerian forces. The first
compilation of such systems revealed that
multi-planet systems are preferentially found among lower-mass
planets \citep{2011ApJ...732L..24L} highlighting the advantages 
of TTVs over radial velocity measurements. Since {\it Kepler}, the search 
for transiting multi-planet systems has seen objects like TRAPPIST-1
\citep{2016Natur.533..221G} with three potential
habitable rocky planets, Kepler-80, a resonant chain of five planets, and 
Kepler-90, the first eight-planet system \citep{2018AJ....155...94S}. 

Transiting multi-planet systems close to resonance allow the
determination of planetary radii {\em and} masses -- and hence bulk
densities -- from transit light curves alone, which has been 
intensively explored by 
\cite{2011ApJS..197....8L}, \cite{2016ApJ...820...39J}, and
\cite{2017AJ....154....5H}. 
A comparison between the two independent mass determinations, namely
using radial velocity and transit timing variations, allows for the 
investigation of systematics due to observational and methodological
biases \citep{2017ApJ...839L...8M}. 

In order to tap the dynamical information of TTVs it is important
to cover a full cycle of orbital momentum and energy exchange between
the planets (henceforth interaction cycle), which can be substantially 
longer than their orbital periods. One of the first lists of systems 
showing TTVs \citep{2013ApJS..208...16M} evidenced the large existing 
fraction of KOIs that could not be used for dynamical analysis due to 
long interaction cycles. These were longer than, or of the order of, {\it Kepler's} 
total observing time. This motivated us to create and coordinate the 
{\it Kepler} Object of Interest Network, 
\citep[KOINet\footnote{\url{koinet.astro.physik.uni-goettingen.de}}][]{vonEssen2017}, 
a network of ground-based telescopes organised to follow up KOIs with 
large amplitude TTVs. The main goal of KOINet is to coordinate already 
existing telescopes to characterise the masses of planets and planetary 
candidates by analysing their observed transit timing variations.

Among the KOINet targets, Kepler-9 is a benchmark system. 
The star is a solar analog and two of its 
planets are close to a 2:1 mean motion resonance, with 
TTV amplitudes of the order of one day. Their deep transits 
($\sim$0.5\,\%) combined with their large interaction times and the 
magnitude of the host star (K$_P$ = 13.803), make this system ideal 
for photometric ground-based follow-up.

The first TTV analysis of the Kepler-9b/c system with an incomplete 
coverage of the interaction cycle had to be complemented with 
(a few) radial velocity measurements \citep{Holman2010} which 
resulted in Saturn-mass planets. The composition of these two
planets was investigated by \citet{Havel2011} from evolutionary models, 
as well as the stellar mass and radius.
Using most or all long-cadence {\it Kepler} data,
several authors revised the planetary masses from TTVs alone 
\citep{Borsato2014,DreizlerOfir2014} finding masses about half 
of the values previously reported in the first paper.
\citet{DreizlerOfir2014} thereby showed that the confirmed innermost 
planet, Kepler-9d, is dynamically independent from this 
near-resonant pair. Recently, a new transit observation for 
Kepler-9b \citep{Wang2017a} was used to correct its transit 
time predictions. Additionally,
the observation of the Rossiter–McLaughlin effect in radial 
velocity measurements of Kepler-9 \citep{Wang2017b} indicates that
the stellar spin axis is very likely aligned with the planetary orbital plane.

In this paper we exploit the large amount of short-cadence {\it Kepler}
data, complemented by long-cadence {\it Kepler} data where short-cadence 
observations are missing, and extended three years in time by adding 
corresponding ground-based light curves from KOINet, all wrapped in 
a detailed photodynamical analysis. The observation of 
the full interaction cycle by the KOINet follow-up, together 
with the comprehensive analysis, results in Kepler-9b and 9c being 
the system with the highest precision planetary mass and radius
determinations. We also constrain the stellar parameters of the
host star and predict the dynamical evolution of the system for 
the next decades.

This paper is divided as follows. We describe the 
new transit observations by the KOINet, their reduction and analysis 
in Sect.~\ref{sec:observations}. The structure of the 
photodynamical model used to analyse KOINet systems is described in 
Sect.~\ref{sec:photdyn}. A description of the results from this 
analysis for the Kepler-9 system can be found in 
Sect.~\ref{sec:results}. These results are discussed in 
Sect.~\ref{sec:discussion}. We end this paper with some conclusions in 
Sect.~\ref{sec:conclusion}.

\section{Observations, data reduction and analysis}
\label{sec:observations}
Between June, 2014 and September 2017 we observed 13 primary transits 
of the Kepler-9b/c planets. The photometric follow-up was carried out 
in the framework of KOINet \citep{vonEssen2017}.  In this work, we combine the 
{\it Kepler} photometry with new, ground-based data which have been collected after 
the nominal time of the {\it Kepler} Space Telescope. This section covers the 
treatment of the new ground-based observations. The photodynamical model described in 
Sect.~\ref{sec:photdyn} was previously fitted to the available {\it Kepler} data
with the aim of getting initial parameters for the ground-based data analysis. A 
description of the photodynamical analysis on the different data sets follows 
in Sect.~\ref{sec:results}. 

\subsection{Data acquisition and main characteristics of the collected photometry}
\begin{table*}[ht!]
\caption{Characteristics of the collected ground-based transit light curves 
of Kepler-9b/c, collected in the framework of KOINet.}  
\label{table:obscond}      
\centering                          
\begin{tabular}{c c c c c c c c}        
\hline\hline
Date		&	Planet	& Telescope	&	$\sigma_\text{res}$	&	N	&	CAD	&	$T_\text{tot}$	& 	TC\\
yyyy.mm.dd 	&		&		&	[ppt]			&	    	&	[sec]	&	[hours]		&	\\
\hline
2014.06.30	&	c	& OLT 1.2m  &	3.6			&	103	    &	79	&	2.3		&	- - - E O \\
2015.06.17	&	b	& ARC 3.5m	&	1.9			&	2075	&	8	&	4.7		&	O I B - - \\
2015.07.25	&	b	& WISE 1m	&	2.2			&	132	    &	166	&	6.1		&	O I B E - \\
            &   b   & CAHA 2.2m &   1.6         &   462     &   57  &   7.4     &   O I B E O \\
	    	&	b	& LIV 2m	&	1.2			&	545	    &	46	&	7.1		&	- - B E O \\
	    	&	b	& NOT 2.5m	&	1.5			&	630	    &	28	&	5.0		&	- - B E O \\
2015.08.14	&	b	& ARC 3.5m	&	2.7			&	2095	&	7	&	3.8		&	- I B E - \\
2015.09.01	&	c	& ARC 3.5m	&	2.2			&   2073	&	4	&	2.5		&	- - B E O \\
2015.10.10	&	b	& IAC 0.8m	&	0.5			&	60	    &	197	&	3.3		&	- I B - - \\
	    	&	b	& TJO 0.8m	&	1.8			&	133	    &	61	&	2.3		&	- I B - - \\
2017.05.17	&	c	& NOT 2.5m	&	0.8			&	219	    &	79	&	4.9		&	O I B E - \\
2017.06.16	&	b	& NOT 2.5m	&	1.1			&	624	    &	30	&	5.4		&	- I B E O \\
2017.06.25	&	c	& NOT 2.5m	&	1.2			&	416	    &	42	&	4.9		&	O I B - - \\
\hline
2014.06.27	&		& OANLH 1m	&\multicolumn{5}{c}{technical difficulties} \\
2014.07.23	&		& OLT 1.2m	&\multicolumn{5}{c}{technical difficulties} \\
2014.08.12	&		& OLT 1.2m	&\multicolumn{5}{c}{weather problems} \\
		    &		& LIV 2m	&\multicolumn{5}{c}{weather problems} \\
2015.03.13	&		& OLT 1.2m	&\multicolumn{5}{c}{weather problems} \\
2016.05.10	&		& ARC 3.5m	&\multicolumn{5}{c}{technical difficulties} \\
2017.09.01	&		& NOT 2.5m	&\multicolumn{5}{c}{weather problems} \\
\hline
\end{tabular}
\tablefoot{The letter code to specify the transit coverage during 
each observation is the following: O: out of transit, before ingress. 
I: ingress. B: flat bottom. E: egress. O: out of transit, after 
egress.}
\end{table*}

Table~\ref{table:obscond} shows the main characteristics of the data 
presented in this work. These are: the date in which the observations 
were performed, in years, months and days; the planet observed during 
transit; an acronym for the telescope used to carry out the 
observations; the precision of the data in parts-per-thousand (ppt); 
the number of frames collected during the night, N; the cadence of the 
data accounting for readout time in seconds, CAD; the total duration 
of the observations in hours, $T_{\rm tot}$; and the transit coverage, TC. 
To increase the photometric precision of the collected data, when 
possible we slightly defocused the telescopes 
\citep{Kjeldsen1992,Southworth2009}. 

A brief description of the main 
characteristics of the telescopes involved in this work follows: 
\begin{itemize}
\item The Oskar L\"uhning Telescope (OLT~1.2m) has a collecting area 
of 1.2~meters and is located at the Hamburger Observatory in Hamburg, 
Germany. The telescope can be used remotely and has a guiding system, 
minimising systematics caused by poor tracking. Although the seeing at 
the observatory is rather poor (typical values are around 
3-4~arcseconds) it stays constant during the night, allowing 
photometric precision in the part-per-thousand level. Here we analyse 
one light curve taken during our first observing season.

\item The Apache Point Observatory hosts the Astrophysical Research 
Consortium 3.5 meter telescope (henceforth ARC~3.5m), and it is 
located in New Mexico, United States of America. Due to the large 
collecting area, with this telescope we collected typically 2000 
observations per observing run. For our observations, the telescope 
was slightly defocused. The photodynamical analysis of Kepler-9 
presented here includes three light curves taken with the ARC~3.5m 
during our second observing campaign, in 2015.

\item The Wise Observatory hosts a 1 meter telescope that is operated 
by Tel Aviv University, Israel (WISE~1m). Here we present one transit 
taken during the second campaign, in 2015.
\item The Centro Astron\'omico Hispano-Alem\'an hosts, among others, a 
2.2~meter telescope (henceforth, CAHA~2.2m). Here we present one 
complete transit observation of Kepler-9b.

\item The 2 meter Liverpool telescope 
\citep[LIV~2m,][]{2004SPIE.5489..679S} is located at the Observatorio 
Roque de los Muchachos, it is fully robotic and is owned and operated 
by Liverpool John Moores University. In this work we present one 
transit observation taken with LIV~2m during our second observing 
season.

\item The Nordic Optical Telescope (NOT~2.5m) is located at the 
Observatorio Roque de los Muchachos in La Palma, Spain. Currently, 
telescope time for KOINet is assigned via a large (3 year) program. 
Here, we analyse four light curves taken between the first and fourth 
observing seasons.

\item The 80 centimetre telescope of the Instituto de Astrof\'isica de 
Canarias (IAC~0.8m) is located at the Observatorio del Teide, in the 
Canary Islands, Spain. Observations were collected by KOINet's 
observers on site. Here we present one light curve taken during our 
second observing season.

\item The Telescopi Joan Or\'{o} (TJO) is a fully robotic 
80~centimetre telescope located at the Observatori Astronomic del 
Montsec, north-east of Spain (henceforth TJO~0.8m). Here we present 
one transit light curve.

\item The Observatorio Astron\'omico Nacional Llano del Hato, 
Venezuela, hosts a 1 meter Zeiss reflector (henceforth 
OANLH~1m). During scheduled observations, the telescope suffered from 
technical difficulties. 
\end{itemize}

\subsection{Data reduction and preparation}
To reduce the impact of our Earth's atmosphere and the associated 
telluric contamination in the I-band, and the absorption of stellar 
light in shorter wavelengths, all of our observations are carried out 
using an R-band filter. Observers always provide a set of calibration 
frames (bias and flat fields) that are used to carry out the 
photometric data reduction. To reduce the data and construct the 
photometric light curves  we use our own IRAF and python-based 
pipelines called {\it Differential Photometry Pipelines for Optimum  
light curves}, DIP$^2$OL. A full description of DIP$^2$OL can be 
found in \cite{vonEssen2017}. In brief, the IRAF component of 
DIP$^2$OL measures fluxes for different reference stars, apertures 
and sky rings; the latter two are set in proportion to
the intra-night averaged full width at half maximum. The python 
counterpart of DIP$^2$OL finds the optimum combination of reference 
stars, aperture, and width of the sky ring that minimises the scatter 
of the photometric light curves. Once the light curves are 
constructed, we transform the time stamps from Universal Time to 
Barycentric Julian Dates in Barycentric Dynamical Time 
(BJD$_\mathrm{TDB}$) using \cite{Eastman2010}'s web tool, all wrapped 
up in a python script.

To detrend the light curves we compute from the photometric science 
frames the time-dependent $x$ and $y$ centroid positions of all the 
stars, the background counts from the sky rings, the integrated flat 
counts for the final aperture centered around the centroid positions, 
the airmass and the seeing. A full description of our detrending 
strategy, how we combine these quantities to construct the detrending 
function, and the extra care in the particular choice and number of 
detrending parameters can be found in Sect.~4.2 of 
\cite{vonEssen2017}. The detrending components, and the time, flux 
and errors, are injected into the transit fitting routine.

\subsection{First data analysis}
\label{subsec:firstanalysis}
Before fitting the full data set using our photodynamical code (see 
Sect.~\ref{sec:photdyn}) we carry out a transit fit to each ground-based light 
curve individually. The main goal of this step is to provide accurate 
error bars for the flux measurements, that are enlarged to account 
for correlated noise \citep[see e.g., ][]{CarterWinn2009}. A 
detailed description of the transit fitting procedure can be found in 
Sect.~4 of \cite{vonEssen2017}. In brief, once the detrending 
components are selected, we fit each transit light curve 
individually. For this we use a detrending times transit  
\citep{MandelAgol2002} model, with a quadratic limb-darkening law and 
thus, quadratic limb-darkening coefficients. The latter are computed 
as described in \cite{vonEssen2013}, for stellar parameters closely 
matching the ones of Kepler-9 \citep{Holman2010} and a Johnson-Cousins R filter 
transmission response. As initial values for all the transit 
parameters, we use the ones given by the photodynamical analysis 
carried out on {\it Kepler} data only. Since the TTVs in this system are so 
large, all of the transit parameters have to be computed for each 
specific light curve. When fitting for the transit parameters, rather 
than using uniform distributions for these parameters, we use 
Gaussian priors with mean and standard deviation equal to the values 
computed from our initial photodynamical analysis on {\it Kepler} data. Only when the transit is 
fully observed do we allow the model to fit for the semi-major axis, the 
inclination, and the planet-to-star radius ratio, along with the 
mid-transit time. Otherwise, all of the transit parameters remain fixed and 
we fit for the mid-transit time only. 

To determine reliable errors for the fitted parameters, we compute 
them from posterior-probability distributions using a Markov chain 
Monte Carlo (MCMC) approach. At this stage, we iterate 100\,000 times 
per transit, and discard a conservative first 20\,\%. Once the best-fit 
parameters are obtained, we compute residual light curves by subtracting 
from the data our best-fit transit-times-detrending models. From the 
residuals we compute the $\beta$ factor as fully described in 
Sect.~4.2 of \cite{vonEssen2017}. Here, we average $\beta$ values 
computed in time bins of 0.8, 0.9, 1, 1.1, and 1.2 times the duration 
of ingress. If this averaged $\beta$ factor is larger than 1, we 
enlarge the photometric error bars by this value, and we repeat the 
MCMC fitting in the exact same fashion as previously explained. The 
{\it raw} light curves obtained after the second MCMC iteration with 
their error bars enlarged, along with the number of detrending 
components per light curve, are the input parameters of the 
photodynamical analysis. As a consistency check, after the 
photodynamical analysis is complete we compare the derived detrending 
coefficients to the ones obtained from individually fitting the light 
curves.

\subsection{Independent check of the timings}
\begin{table}[ht!]
    \centering
    \caption{\label{tab:tos_simul}{\it Top:} Transit parameters obtained 
    fitting one light curve of Kepler-9b observed with CAHA~2.2m on the 
    night of 2015.07.25. From left to right: transit parameter (TP), 
    transit fitted parameter (TFP) along with 1-$\sigma$ uncertainties, 
    and value predicted by our photodynamical analysis (PDA) applied at 
    this stage to {\it Kepler} data only. {\it Bottom:} Mid-transit times 
    obtained fitting the three remaining incomplete transit light curves. 
    The first column shows the acronym for the telescope, and the second 
    column the timings along with 1-$\sigma$ uncertainties.}
    \begin{tabular}{l c c}
    \hline\hline
    TP, CAHA 2.2m &              TFP  & PDA \\
    \hline
    $a/R_S$        & 29.3 $\pm$ 0.2    &  29.27 $\pm$ 0.1 \\
    $i\;[^{\circ}]$ & 88.77 $\pm$ 0.03  &  88.76 $\pm$ 0.2 \\
    $R_P/R_S$    & 0.085 $\pm$ 0.001 &  0.079 $\pm$ 0.003 \\
    \hline
    Telescope      &   $T_0$ $\pm$ 1-$\sigma$ & \\
    \hline
    CAHA 2.2m,     & 229.4598 $\pm$ 0.0008 & 229.4606 $\pm$ 0.005 \\    
    LIV 2m         & 229.4629 $\pm$ 0.0009 & \\
    WISE 1m        & 229.4623 $\pm$ 0.0026 & \\
    NOT 2.5m       & 229.4582 $\pm$ 0.0018 & \\
    \hline
    \end{tabular}
    \tablefoot{$T_0$'s are given in BJD$_{\mathrm{TDB}}$ - 2457000.}
\end{table}
\begin{figure}[ht!]
\resizebox{\hsize}{!}{\includegraphics{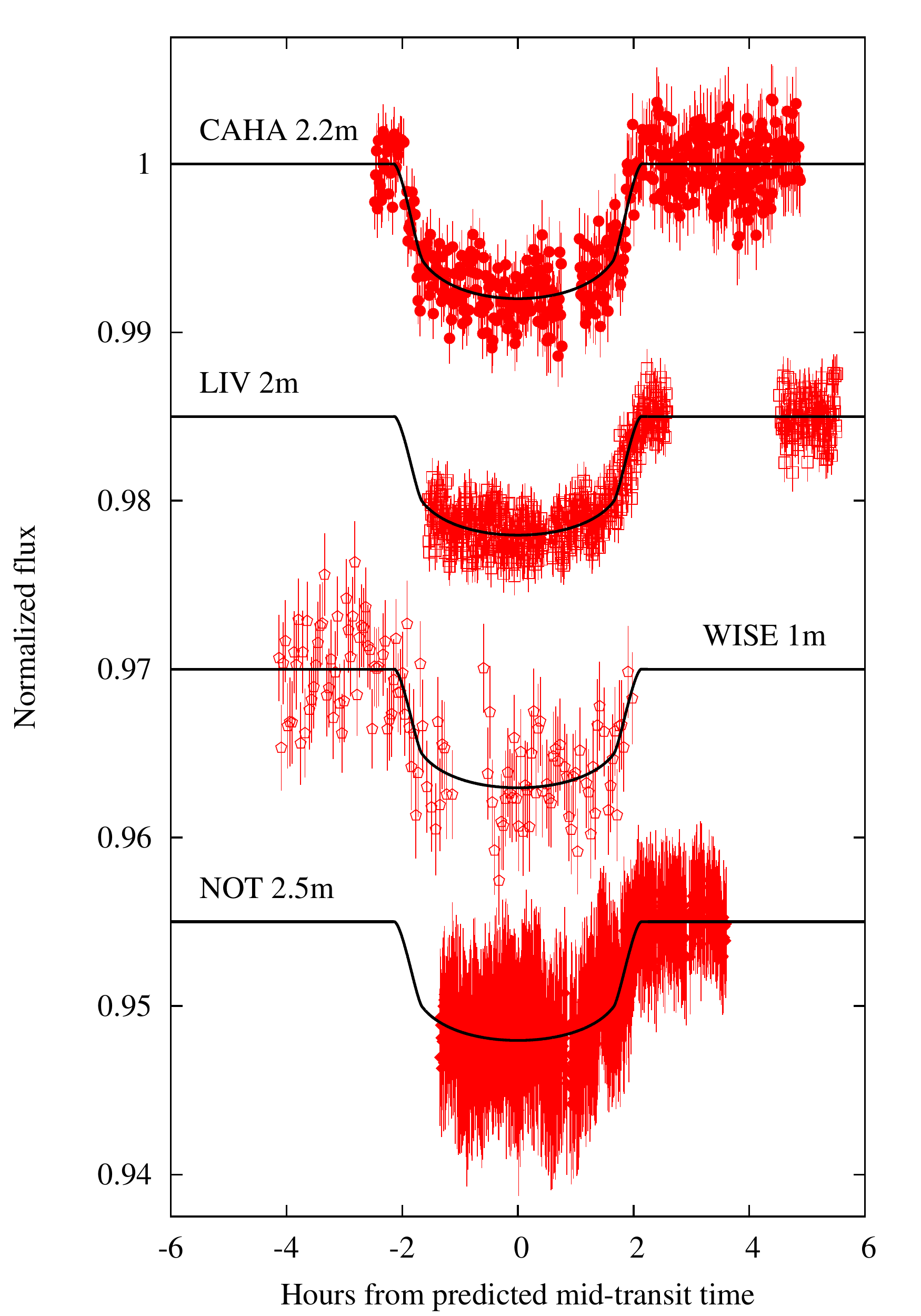}}
\caption{Detrended transits of Kepler-9b observed on 
July 25, 2015, by four different telescopes. The transits are 
artificially shifted for better visual inspection, and plotted as a 
function of hours from the predicted mid-transit time to appreciate 
the duration of the observations. Each light curve has been labelled 
according to the corresponding telescope.}
\label{fig:gbd}
\end{figure}

The use of KOINet to carry out TTV studies relates observations taken 
with several telescopes. As a consequence, the timings will be subject 
to the accuracy of the ground-based observatories, and the success of 
KOINet will rely on how accurately the many observatories involved in 
our photometric follow-up know their own times. 

In order to investigate this, on the night of July 25, 2015 we 
observed Kepler-9b using four different telescopes, namely CAHA~2.2m, 
LIV~2m, WISE~1m, and NOT~2.5m. Only in the case of CAHA 2.2m did we have 
full transit coverage. After fitting for the transit parameters as 
previously specified, we obtained in this case the semi-major axis, 
$a/R_S$, the inclination, $i$, the planet-to-star radii ratio, 
$R_P/R_S$, and the mid-transit time, $T_0$. The derived values, along 
with their 1-$\sigma$ uncertainties, can be found in the top part of 
Table~\ref{tab:tos_simul}. Within errors all the fitted parameters 
are consistent with the values predicted by our photodynamical 
analysis. The bottom part of the same table shows the individual 
mid-transit times obtained from fitting all the transit parameters 
for CAHA~2.2m, and fixing all values except the mid-transit times for 
the remaining three. All mid-transit times are mutually consistent.

Figure~\ref{fig:gbd} shows the quality of our reduction and analysis 
procedure. From top to bottom, the light curves of Kepler-9b obtained 
with CAHA~2.2m in filled circles, LIV~2m in empty squares, WISE~1m in 
empty polygons, and NOT~2.5m in filled diamonds are shown. The light 
curves have been shifted to the predicted mid-transit time. A visual 
inspection confirms the equivalency of the derived 
mid-transit times. The consistency among mid-transit times alleviates the 
uncertainty that exists when using different sites to follow-up 
one target.

\section{The photodynamical model}
\label{sec:photdyn}
With the aim of producing a tool to determine planetary masses that is 
applicable to all of our KOINet objects, we developed a photodynamical model. 
Our light curve analysis is based on an n-body simulation of the 
planetary system over the time span of the observations, combined 
with a transit light curve model. The numerical integration is 
implemented in the \texttt{Mercury6} package by 
\citet{Chambers1999}. We use the second-order mixed-variable 
symplectic (MVS) algorithm of the package, which is faster than 
the Bulirsch-Stoer (BS) algorithm but still applicable. 
The BS integrator would have two advantages, that are the possibility of 
simulating close encounters and the precision in high frequency terms of the 
Hamiltonian \citep[discussed by][]{Deck2014}. The former is insignificant as 
the Kepler-9b/c system does not perform close encounters. The latter was tested 
to be negligible in our analysis. We calculated the difference of the same TTV 
model derived with the MVS integrator and the BS algorithm. Within a 50 year 
integration the difference shows a small slope, that can be corrected by a small 
change in the mean period smaller than $0.5$\;s and an oscillation with increasing 
amplitude. The amplitude of the oscillation is in maximum (in this 50 years) of the 
order of $55$\;s which is the order of the precision in the TTVs. For the 8 years of 
Kepler-9 observations the MVS integrator has a sufficient precision.
We added a first order 
post-Newtonian correction \citep{1995PhRvD..52..821K} and wrote 
a python-wrapper for \texttt{Mercury6} (Husser, priv. comm.). 
From the n-body simulation we extract the projected 
distance between planet and star centres, that are in turn used to 
calculate the transit light curve through the \citet{MandelAgol2002} 
model. Here we use a quadratic limb darkening law already 
implemented in the \texttt{occultquad} routine. Finally, for each 
individual planet we correct the output time by the light-travel time 
effect.

As the numerical integration and its processing is computationally 
time consuming, we first carry out a coarse integration with a step 
size equal to only a twentieth of the period of the innermost 
simulated planet. A more detailed integration is produced for the 
parts where transits take place and where data are available. For 
these cases, the detailed integration is done with a step size of 
$0.01$\;days, which corresponds to a light curve accuracy of 
$0.01$\;ppm for long cadence \textit{Kepler}-data. 
This accuracy is measured by the mean difference of transit 
light curves between a model with the given step size and a model 
with half the step size.
For transit light 
curves, a time step comparable to ingress/egress duration would have a 
significant impact in the derivation of the transit parameters 
\citep{Kipping2010}. Therefore, we calculate the transit model on a 
fine grid ($\sim$1 minute, when needed) and we rebin it to the actual 
data points. We describe this in more detail in 
Sect.~\ref{sec:results}. For our model calculations, we define the 
$x-y$-plane as the plane of the sky, with its origin placed at the 
stellar centre. Therefore, these coordinates coincide with the 
projected distances between planet and star mid points. The positive 
$z$-axis corresponds to the line of sight, so that the planets transit 
with positive $z$-values. The sampling of the \texttt{Mercury6} 
integration does not match the observation times. To interpolate the 
projected distances from the \texttt{Mercury6} results, we model them 
with a hyperbola in the range of a transit. The 
\citet{MandelAgol2002} model is calculated for the observation points 
by these interpolated projected distances, quadratic limb darkening 
coefficients, and the planet-star radii ratios.

To explore the parameter space, our model is coupled to the Markov 
chain Monte Carlo \texttt{emcee} algorithm \citep{emcee} accessible 
in the \texttt{PyAstronomy\footnote{\url{https://github.com/sczesla/PyAstronomy}}} 
library. All fitting parameters have uniform priors with large limits 
with the only purpose of avoiding non-physical results. Our choice of 
free parameters is guided by the modelling rather than by the 
observations. For instance, \texttt{Mercury6} uses the semi-major 
axis, $a$ of the planet as input value. Instead of the period, $P$, 
we therefore use a correction factor to a mean semi-major axis 
$a_\text{adjust}$ as a free parameter. The mean semi-major axis is 
calculated through Kepler's third law from the mean period of the 
transit times of the planets. In addition, the mean anomaly, $M$, 
is calculated from this mean period, as well as the reference 
time, $\Delta T_0$. As a free parameter we have an addend to 
this derived mean anomaly $M_\text{adjust}$.
Furthermore, \texttt{Mercury6} uses the eccentricity, $e$ and the 
three angles that describe the position of the orbits on the sky. They 
are the orbital inclination, $i$, the argument of the periastron, 
$\omega$, and the longitude of the ascending node, $\Omega$. As the 
orientation in the plane of the sky is not directly measurable, 
$\Omega$ is fixed to zero for the innermost simulated planet. In this 
way, the corresponding values of the other planets show the 
difference in comparison to this first one. Last but not least, 
\texttt{Mercury6} requires the masses, $m$, of the central star and 
the planets. These are given by an absolute value for the central 
star, the ratio of the masses of the innermost simulated planet to 
the central star, and the ratio of masses of the other planets to the 
innermost planet.

In order to calculate the transit light curve from the output of 
\texttt{Mercury6}, the stellar radius, $R_S$, is required to 
calculate the relative planet-star distance normalised to the stellar 
radius. The transit measurements constrain the stellar density 
\citep{AgolFabrycky2017}, but we choose to directly use the required 
model parameters. Instead of the stellar density we input the stellar 
mass and radius, but fixing one of them during the modelling. In 
addition, the \texttt{occultquad} routine requires the planet-star 
radius ratio, $R_p/R_S$, and the two quadratic limb darkening 
coefficients, $c_1$ and $c_2$.

\section{Dynamical analysis of Kepler-9}
\label{sec:results}
To dynamically characterise the Kepler-9b/c system we carried out 
three different approaches. Firstly, in order to compare the 
photodynamical model with the dynamical analysis of only transit 
times, we fitted our model to quarter 1 through 16 {\it Kepler} long 
cadence data (hereafter data set I). This allowed us to compare our 
results to the ones given by \citet{DreizlerOfir2014}. Secondly, we 
attempted to constrain the stellar radius by means of {\it Kepler} short 
cadence data, since they have a thirty times higher sampling. Towards 
this end, we replaced {\it Kepler} long cadence with short cadence data 
when available. Specifically for Kepler-9, short cadence data is 
available between quarter 7 and 17 (data set II). Finally, the model 
is applied to the full data set, which comprises long cadence data 
for {\it Kepler} quarters 1 to 6, short cadence data covering quarters 7 to 
17, and all new ground-based light curves, 13 in total, that were 
collected via the KOINet (data set III).

The results from all the data analysis and a comparison to previous 
analysis are listed in Table~\ref{table:results}. The top part of the 
table shows the stellar parameters. The literature values of the 
stellar radius and density parameters are taken from 
\citet{Havel2011}, and the respective quadratic limb darkening values 
are taken from the \texttt{NASA Exoplanet Archive} \citep{NASAexopA}.
The bottom part of Table~\ref{table:results} shows the derived 
planetary parameters. They are compared to the results given by 
\citet{DreizlerOfir2014}. In this work the authors modelled the 
transits observed in long cadence data individually, from where the 
mid-transit times were derived. Afterwards, they dynamically modelled 
these transit times. 

\begin{table*}
\caption{Stellar and planetary parameters derived from the photodynamical 
modelling of data set I in the second column, data set II in the third 
column, data set III in the fourth column, along with bibliographic 
values \citep{DreizlerOfir2014} in the fifth column for comparison.
Given are the median and standard deviation values from the MCMC
posterior distributions.
For the stellar radius and density, the bibliographic values are taken 
from \citet{Havel2011}. The quadratic limb darkening coefficients are 
taken from \citet{NASAexopA}. The sixth column displays some parameters 
corrected by investigating stellar evolution models in 
section~\ref{subsec:stellarevol}. The osculating orbital elements are 
given at a reference time, $\text{BJD}=2454933.0$.}             
\label{table:results}      
\centering                          
\begin{tabular}{c c c c c c}        
\hline\hline															
Parameter		&	Data set I 	&	Data set II  & 	Data set III	&	Literature	&	MESA	\\
 \hline
\multicolumn{2}{l}{Stellar parameters:} & & &  & \\
\hline															
$m_S\;[M_\sun]$ &$1.05(3)$ (fixed) &$1.05(3)$ (fixed) &$1.05(3)$ (fixed) &$1.05(3)$ & $1.04_{-0.04}^{+0.07}$\\
$R_S\;[R_\sun]$	&$	0.947	(	21	)$&$	0.9755	(	92	)$&$	0.9742	(	83	)$&$	1.05	(	6	)$  & $0.971_{-0.021}^{+0.030}$  \\
$\rho_S^* [\text{g\,cm}^{-3}]$	&$	1.74	(   12	)$&$	1.596	(   45	)$&$	1.603	(   41  )$&$	1.12	(	27	)$  &   \\
$\rho_S^* [\text{g\,cm}^{-3}]$ ($\sigma_{m_S}$ prop.)	&$	1.74	(	13	)$&$	1.596	(	64	)$&$	1.603	(	61	)$&$	1.12	(	27	)$  &   \\
$c_{1,\text{\it Kepler}}$	&$	0.281	(	53	)$&$	0.361	(	51	)$&$	0.351	(	47	)$&$	0.4089		$  &   \\	
$c_{2,\text{\it Kepler}}$	&$	0.410	(	95	)$&$	0.251	(	78	)$&$	0.269	(	71	)$&$	0.2623		$  &   \\	
\hline																	
\multicolumn{2}{l}{Planetary parameters:} & & &  &   \\																	
\hline																	
$m_b/m_S$	&$	0.0001271	(	11	)$&$	0.0001271	(	11	)$&$	0.0001281	(	11	)$&$	0.000129	(	2	)$  &   \\
$m_c/m_b$	&$	0.68911	(	26	)$&$	0.68846	(	22	)$&$	0.68849	(	20	)$&$	0.6875	(	3	)$  &   \\
$m_b^*\;[M_\Earth]$	&$	44.36	(	44	)$&$	44.51	(	32	)$&$	44.71	(	24	)$&$	45.1	(	15	)$  &   \\
$m_c^* \;[M_\Earth]$	&$	30.57	(	30	)$&$	30.64	(	22	)$&$	30.79	(	17	)$&$	31	(	1	)$  &   \\
$m_b^* \;[M_\Earth]$ ($\sigma_{m_S}$ prop.)	&$	44.4	(	13	)$&$	44.5	(	13	)$&$	44.7	(	13	)$&$	45.1	(	15	)$  & $44.4_{-1.7}^{+3.0}$  \\
$m_c^* \;[M_\Earth]$ ($\sigma_{m_S}$ prop.)	&$	30.57	(	92	)$&$	30.64	(	90	)$&$	30.79	(	90	)$&$	31	(	1	)$  &  $30.5_{-1.2}^{+2.1}$ \\
$a_{b,\text{adjust}}\;[\text{AU}]$	&$	0.9992801	(	21	)$&$	0.9992811	(	11	)$&$	0.9992801	(	11	)$&		-	  &   \\	
$a_{c,\text{adjust}}\;[\text{AU}]$	&$	1.0015531	(	31	)$&$	1.0015521	(	31	)$&$	1.0015531	(	21	)$&		-	  &   \\	
$a_b^* \;[\text{AU}]$	&$	0.14276083	(	21	)$&$	0.14276096	(	16	)$&$	0.14276088	(	14	)$&$	0.143	(	1	)$  &   \\
$a_c^* \;[\text{AU}]$	&$	0.22889883	(	83	)$&$	0.22889869	(	63	)$&$	0.22889876	(	53	)$&$	0.229	(	2	)$  &   \\
$a_b^* \;[\text{AU}]$ ($\sigma_{m_S}$ prop.)	&$	0.1428	(	14	)$&$	0.1428	(	14	)$&$	0.1428	(	14	)$&$	0.143	(	1	)$  &  $0.1423_{-0.0018}^{+0.0032}$ \\
$a_c^* \;[\text{AU}]$ ($\sigma_{m_S}$ prop.)	&$	0.2289	(	22	)$&$	0.2289	(	22	)$&$	0.2289	(	22	)$&$	0.229	(	2	)$  &  $0.2282_{-0.0029}^{+0.0051}$ \\
$e_b$	&$	0.06437	(	74	)$&$	0.06412	(	54	)$&$	0.06378	(	40	)$&$	0.063	(	1	)$  &   \\
$e_c$	&$	0.068026	(	92	)$&$	0.067974	(	73	)$&$	0.067990	(	68	)$&$	0.0684	(	2	)$  &   \\
$i_b\;[^\degree]$	&$	89.037	(	85	)$&$	88.931	(	33	)$&$	88.936	(	30	)$&$	87.1	(	7	)$  &   \\
$i_c\;[^\degree]$	&$	89.229	(	41	)$&$	89.177	(	17	)$&$	89.180	(	15	)$&$	87.2	(	7	)$  &   \\
$\omega_b\;[^\degree]$	&$	357.17	(	33	)$&$	357.10	(	24	)$&$	356.98	(	20	)$&$	356.9	(	5	)$  &   \\
$\omega_c\;[^\degree]$	&$	169.29	(	11	)$&$	169.215	(	95	)$&$	169.194	(	73	)$&$	169.3	(	2	)$  &   \\
$M_{b,\text{adjust}}\;[^\degree]$	&$	4.0426	(	48	)$&$	4.0441	(	50	)$&$	4.0459	(	39	)$&		-	  &   \\	
$M_{c,\text{adjust}}\;[^\degree]$	&$	-3.2629	(	63	)$&$	-3.2654	(	53	)$&$	-3.2648	(	46	)$&		-	  &   \\	
$M_b^* \;[^\degree]$	&$	337.01	(	41	)$&$	337.12	(	30	)$&$	337.28	(	24	)$&$	337.4	(	6	)$  &   \\
$M_c^* \;[^\degree]$	&$	313.489	(	97	)$&$	313.553	(	87	)$&$	313.575	(	67	)$&$	313.5	(	1	)$  &   \\
$\Omega_b\;[^\degree]$ & $0$ (fixed) & $0$ (fixed) & $0$ (fixed) & $0$ (fixed)   &   \\																	
$\Omega_c\;[^\degree]$	&$	-1.37	(	15	)$&$	-1.244	(	88	)$&$	-1.268	(	75	)$&$	0	$(fixed)  &   \\		
$R_b/R_S$	&$	0.07644	(	60	)$&$	0.07766	(	30	)$&$	0.07759	(	27	)$&$	0.0825	(	1	)$  &   \\
$R_c/R_S$	&$	0.07498	(	60	)$&$	0.07601	(	32	)$&$	0.07595	(	28	)$&$	0.0796	(	2	)$  &   \\
$R_b^* \;[R_\Earth]$	&$	7.91	(	24	)$&$	8.27	(	11	)$&$	8.252	(	94	)$&$	11.1	(	1	)$  &  $8.22_{-0.18}^{+0.26}$ \\
$R_c^* \;[R_\Earth]$	&$	7.76	(	23	)$&$	8.10	(	10	)$&$	8.077	(	92	)$&$	10.7	(	1	)$  &  $8.05_{-0.18}^{+0.25}$ \\
$\rho_b^* [\text{g\;cm}^{-3}]$	&$	0.495   (	47	)$&$	0.434	(	17	)$&$	0.439	(	15	)$&$	0.18	(	1	)$  &   \\
$\rho_c^* [\text{g\;cm}^{-3}]$	&$	0.362   (	34	)$&$	0.319	(	12	)$&$	0.322	(	11	)$&$	0.14	(	1	)$  &   \\
$\rho_b^* [\text{g\;cm}^{-3}]$ ($\sigma_{m_S}$ prop.)	&$	0.495	(	60	)$&$	0.434	(	24	)$&$	0.439	(	23	)$&$	0.18	(	1	)$  &   \\
$\rho_c^* [\text{g\;cm}^{-3}]$ ($\sigma_{m_S}$ prop.)	&$	0.362	(	44	)$&$	0.319	(	18	)$&$	0.322	(	17	)$&$	0.14	(	1	)$  &   \\
\hline															
\end{tabular}
\tablefoot{$^*$ derived, not fitted parameters}
\end{table*}

The osculating orbital elements are given at a reference 
time, $\text{BJD}=2454933.0$. Fitting the transit times found in 
\citet{DreizlerOfir2014} with a linear time-dependent model we 
obtained the reference times, $\Delta T_b=25.26\,\text{d}$ and 
$\Delta T_c=-3.08\,\text{d}$ as the intercepts, and the mean periods, 
$P_b=19.247\,\text{d}$ and $P_c=38.944\,\text{d}$ as slopes. The 
reference times and mean periods are used for the determination of 
the semi major axis and the mean anomaly for all data sets, as 
described previously in Sect.~\ref{sec:observations}.

During our photodynamical modelling we chose to fix the stellar mass 
to its literature value, $m_S=1.05\pm0.03\,M_\sun$ \citep{Havel2011}. 
Derived parameters that depend on this value are the planetary 
masses, as the model parameters are given with respect to the stellar 
mass. Therefore, the uncertainties of the derived parameters are 
increased using error propagation including the uncertainty of the 
stellar mass, $\sigma_{m_S} = 0.03\,M_\sun$. When applied, in 
Table~\ref{table:results} these parameters are labelled with 
"$\sigma_{m_S}$ prop.". The calculated densities of the star 
and the planets depend on the stellar mass in the same way.
The semi-major axes are also affected. 
These are computed from the mean period through Kepler's third law, 
which also includes the stellar and planetary masses. As a consequence, 
this error is also propagated into the uncertainty of the semi-major 
axis. 

A quick comparative look at Table~\ref{table:results} shows how the 
limb darkening coefficients obtained modelling data set I 
significantly differ from their literature values. We address this 
issue in Sect.~\ref{sec:discussion}. With this exception, within 
1-$\sigma$ errors all planetary parameters are in agreement with 
prior results. The errorbars decrease from modelling data set I to 
III. The causes of this will be given in detail in the following 
sections.

\subsection{Treatment of the Kepler data}
To prepare {\it Kepler's} transit photometry we first extracted three times 
the transit duration symmetrically around each transit mid point. To 
account for intrinsic stellar photometric variability we normalised 
each transit light curve dividing this by a time-dependent second 
order polynomial fitted to the out-of-transit data points. To obtain 
the coefficients of the polynomial functions, we used a simple least 
squares minimisation routine. As previously mentioned, for long 
cadence data the light curve model is oversampled by a factor of 30 
and rebinned to the actual data points. This procedure is not 
necessary for short cadence data. The high signal-to-noise ratio of 
{\it Kepler} data allowed us to include the quadratic limb darkening 
coefficients into our model budget.

\subsection{Treatment of ground-based data}
Due to the lower signal-to-noise ratio of the ground-based data, we 
fixed the quadratic limb darkening coefficients to values derived 
from stellar evolution models for the R-band filter, which we used 
for all our observations. For stellar parameters closely matching the 
ones of Kepler-9, the derived limb darkening coefficients are 
$c_1=0.46$ and $c_2=0.17$. The best matching coefficients of the 
detrending components for each ground-based observation, that are 
derived previously (see Sect.~\ref{subsec:firstanalysis}), are 
calculated as a linear combination at each call of the photodynamical 
model.

\subsection{Statistical considerations}
We performed the analysis of data set I with 36 walkers each, 
iterating over 30\,000 steps. The starting parameters of the walkers 
are randomly chosen from a normal distribution around the parameter 
results of \citet{DreizlerOfir2014} with a 3-$\sigma$ width. The 
walkers needed 2\,000 iterations to burn in, with the exception of one that 
finished in a higher $\chi^2$ minimum. Thus, our results are derived 
from 35 walkers with 28\,000 iterations each. 
We calculated the autocorrelation time for each parameter following
\citet{GoodmanWeare2010}, but averaging over the autocorrelation function
per walker instead of averaging directly over the walker values like discussed
in the Blog by Dan Foreman-Mackey\footnote{\url{http://dfm.io/posts/autocorr}}.
These calculations results in an autocorrelation time of 1\,853 in average
(2\,771 in maximum), which gives us an effective sample size of 528 (353 in minimum).
Each 
parameter shows a Gaussian posterior distribution from which we 
extract the median and standard deviation values as best-fit values 
and errors, respectively. Our results are shown in 
Table~\ref{table:results}. The best-fit solution has a reduced 
$\chi^2$ of $1.48$.

The analysis of data set II is performed using 36 walkers with 
20\,000 iterations each. In this case, after 4\,000 iterations they 
burned in, with the exception of 2 walkers that ended in a higher 
$\chi^2$ minimum. 
The autocorrelation time averages out at 927 (1\,648 in maximum), which
gives an effective sample size of 586 (330 in minimum).
The resulting parameters are derived using the 
median and standard deviation of the posterior Gaussian distribution.
The best solution of this analysis has a 
reduced $\chi^2$ of $1.06$.

The modelling of data set III is accomplished by 36 
walkers with 20\,000 iterations each. Thirty-five of the walkers burned in 
after 2\,000 iterations. The resulting Gaussian distributions of 
the 630\,000 iterations for the parameters and their correlations 
can be seen in the appendix in Fig.~\ref{fig:corner_mass} 
for the mass-dependent parameters, in Fig.~\ref{fig:corner_radius} 
for the radius-dependent parameters, and in total in 
Fig.~\ref{fig:corner}. Our best-fit solution has a reduced $\chi^2$ 
of $0.97$. The autocorrelation length of this analysis is given by a 
value of 694 in average (1105 in maximum). This results in an effective 
sample size of 907 (570 in minimum).

\subsection{Results}
\begin{figure}
\resizebox{\hsize}{!}{\includegraphics{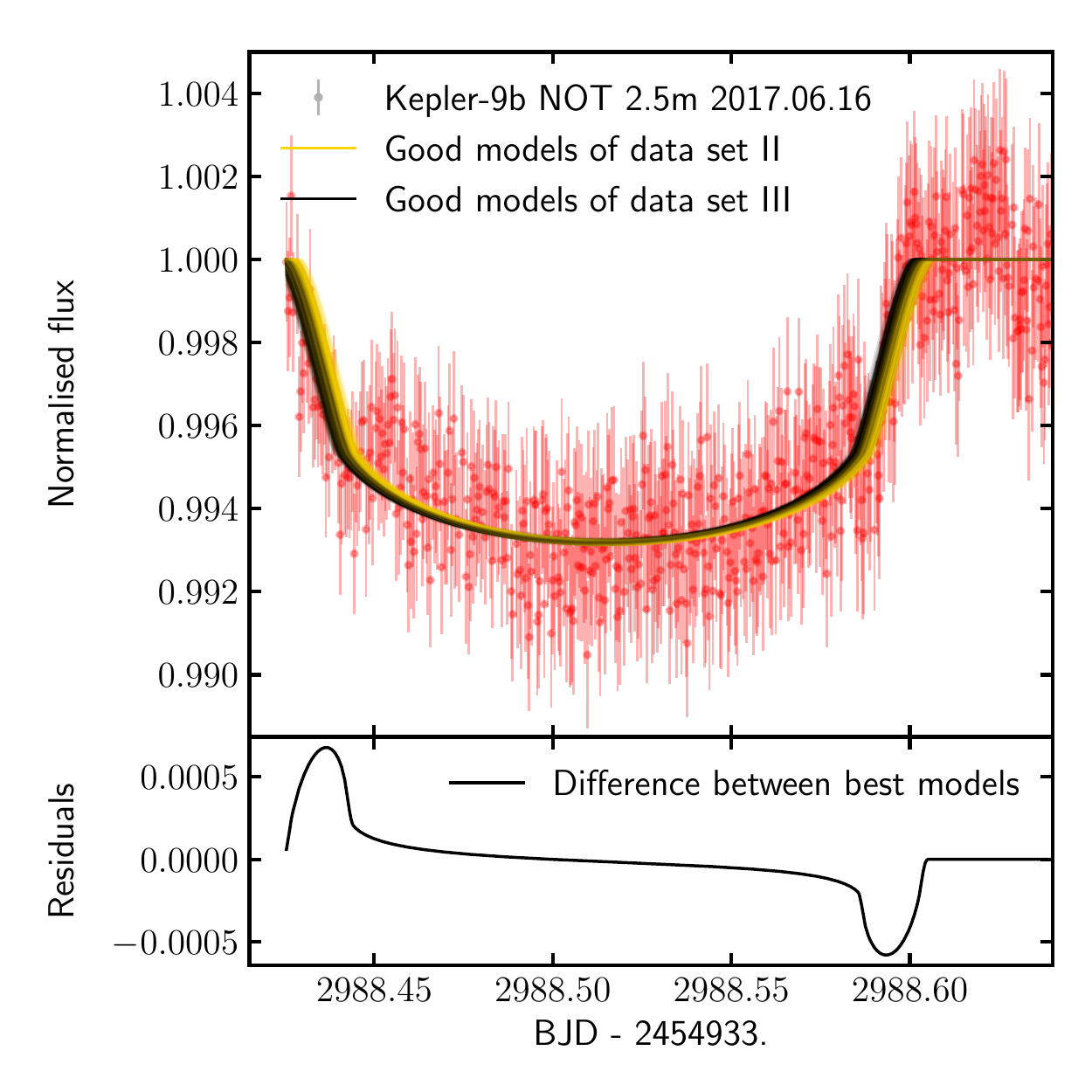}}
\resizebox{\hsize}{!}{\includegraphics{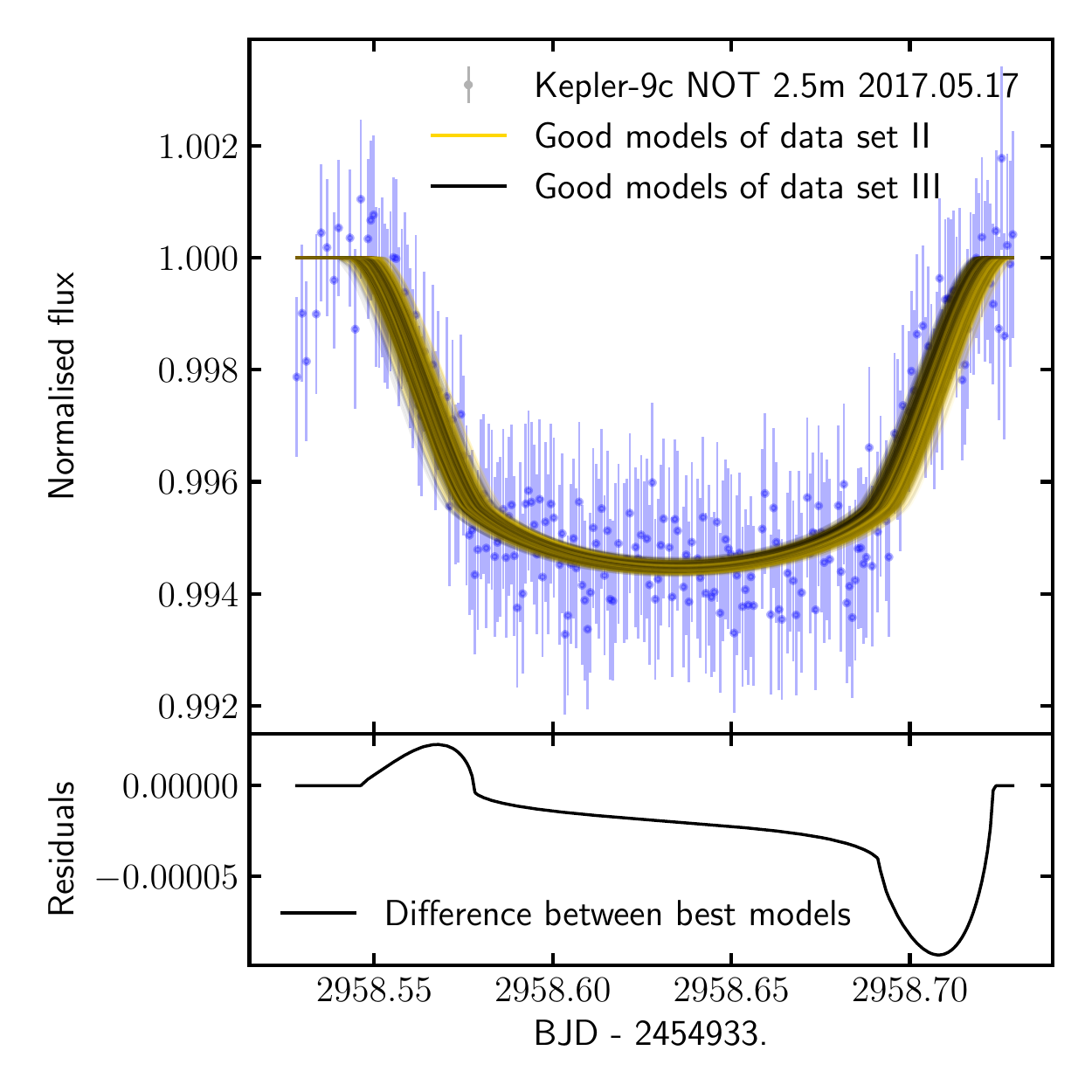}}
\caption{Examples of new obtained transit light curves for Kepler-9b 
in red (top), observed on June 16, 2017, and Kepler-9c 
in blue (bottom), observed on May 17, 2017. Both 
transits were observed using the NOT~2.5m telescope. Overplotted is 
the variation of 500 randomly chosen good models by modelling data set II 
in yellow and modelling of the full data set (data set III) in black. 
The residuals plot shows the difference between the best models of these 
two data sets.}
\label{fig:newtransits}
\end{figure}
\begin{figure*}
\resizebox{\hsize}{!}{\includegraphics{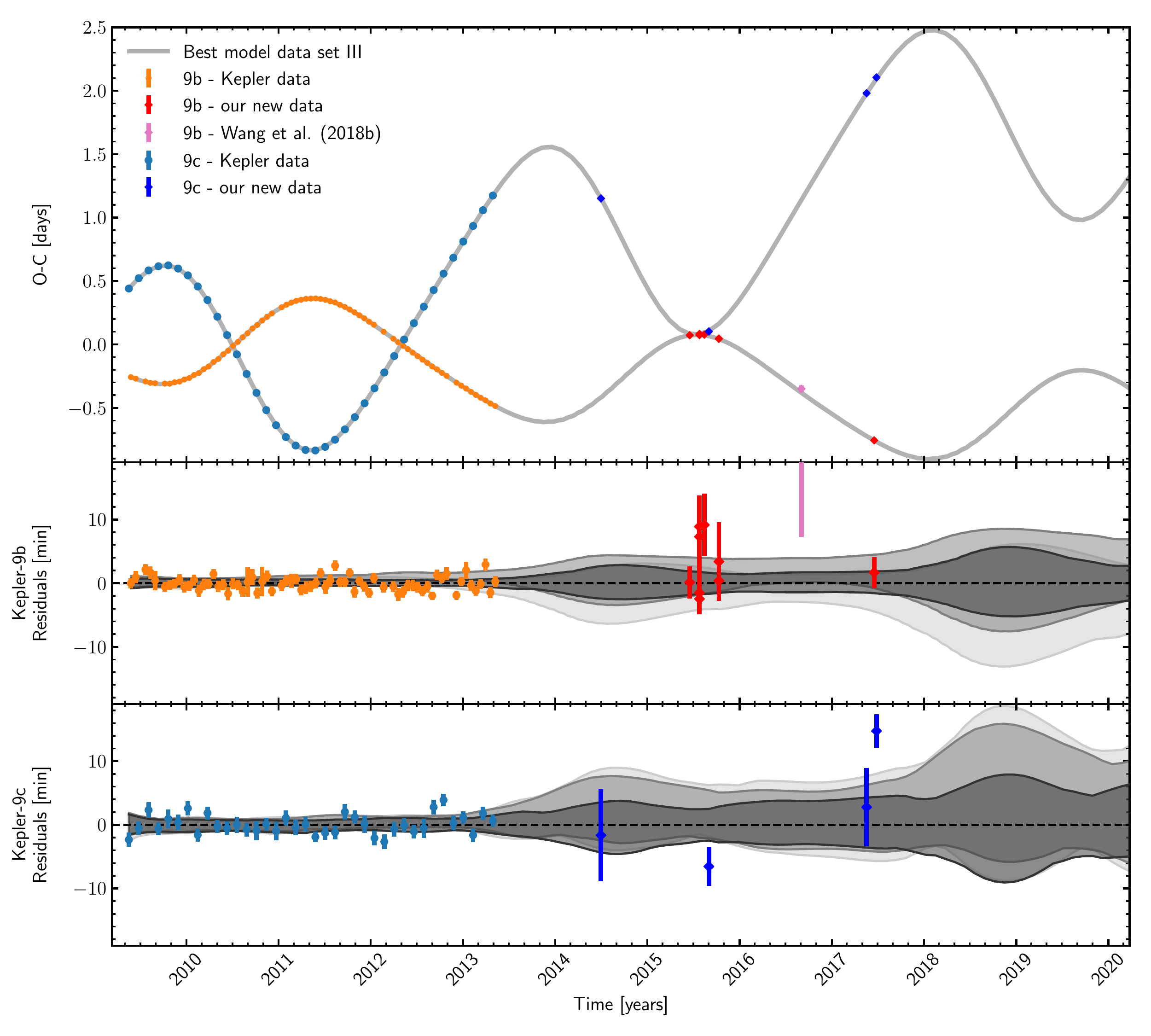}}
\caption{The O-C diagram of transit times from photodynamical 
modelling of the data set III and the predictions until 2020 (top). 
Calculated (C) are transit times from a linear ephemeris modelled at 
the transit times found by \citet{DreizlerOfir2014}. The residual plots
(middle: Kepler-9b, bottom: Kepler-9c) shows the 99.74\% confidence 
interval of 1000 randomly chosen good models in comparison the best 
found model of data set 3. From light to dark grey: modelling of 
data set I, II and III.
The {\it Kepler} transit times are derived by single 
transit modelling. The new transit time data points originate from 
the first analysis described in Sect.~\ref{subsec:firstanalysis}.}
\label{fig:ttv}
\end{figure*}
The comparison of the best models to the most recent light curves 
from 2017 displayed in Fig.~\ref{fig:newtransits} clearly shows how 
the inclusion of our new ground-based light curves leads to the 
improvement of the derived parameters. The top plot shows a Kepler-9b 
transit light curve in red observed on June 16, 2017. 
The bottom  plot shows a Kepler-9c transit light curve in blue 
observed on May 17, 2017. Both light curves were 
obtained using the NOT~2.5m telescope. The variation of 500 randomly 
chosen good models for data set II are given by the light transparent 
yellow areas, which can be compared to the corresponding ones 
obtained including all new ground-based data (data set III). These 
are plotted in the figures with a light transparent black area. 
Additionally, the difference between the best model of data set II 
and III can be seen in the bottom panels of the plots (henceforth 
residual plots). Comparing the yellow and the black areas shows a 
slight narrowing of the model variation for data set III, which is 
reflected by the slightly smaller errorbars in 
Table~\ref{table:results}. In the case of Kepler-9b, the transit 
models slightly shift towards earlier transits when all ground-based 
data are included. This can be recognised by comparing the yellow and 
black areas, but it is more obvious in the residual plots. 
A larger change between modelling the different data sets appears for 
Kepler-9c. The residuals of the best model for this transit shows an
asymmetric difference between the modelling of data set II and III. 
That means an adjustment not only in the transit time, but also in 
the transit shape. 

The obtained detrended ground-based transit light curves are shown in 
Fig.~\ref{fig:allnewtransits}, together with the best 
photodynamical model in grey, and the variation of 500 randomly chosen 
good fitting models in black. The data corresponding to Kepler-9b are 
plotted in red, and the ones of Kepler-9c are plotted in blue. Each 
observation has its own sub-figure, where the date and the used telescope 
are indicated. The transits that were observed from 
different sites simultaneously are artificially shifted to allow 
for a visual inspection. Raw photometry are available for download. 

Another derivable parameter of our photodynamical model are the 
transit times. Figure~\ref{fig:ttv} shows the O-C diagram of the 
transit times measured by individually fitting the {\it Kepler} data, as 
well as the newly obtained ground-based data, in comparison to the 
results of modelling data set III. Also included is the mid transit 
time of Kepler-9b obtained by \citet{Wang2017a}, about $2\sigma$ off from 
our model and our new data. Unfortunately, the photometric data are 
not published so we could not include it in our photodynamic
analysis. The top part of Fig.~\ref{fig:ttv} shows the O-C diagram 
with the transit times from {\it Kepler} data in orange for Kepler-9b and 
in light blue for Kepler-9c. The O-C data from the new KOINet 
observations are shown in red for Kepler-9b and blue for Kepler-9c. 
The mid-time derived by \citet{Wang2017a} is shown in pink. The 
transit times from the best photodynamical model of data set III 
minus the linear trend are presented as grey lines. The middle part 
of this figure shows the residuals for Kepler-9b with the same colour 
identification, and the residuals of Kepler-9c are shown respectively 
in the bottom part. In both residual plots, 
the 99.74\% confidence interval of 1000 randomly chosen good models 
of the different data sets in comparison to the best model of 
data set 3 are plotted as grey areas. The light grey area belongs to 
the modelling of data set I, middle grey to data set II and
dark grey to data set III respectively. The differences in the amplitude
of the variations of the models compared to the best model are discussed 
in the next section. 
In Table~\ref{table:ttv} we provide transit time predictions 
from modelling data set III for the next ten years.

\section{Discussion}
\label{sec:discussion}
The results of the photodynamical modelling of Kepler-9b/c require 
some interpretation. In this section we will first discuss the 
dynamical stability of the derived system model, followed by 
a discussion of the transit timing variations along with their 
prediction for future observations. Furthermore, an emphasis will be 
given to the transit shape variations and the consequential
prediction of disappearing transits for Kepler-9c. Moreover, we will 
address the stellar activity and, connected to this, we investigate 
the stellar mass, radius, and age. The age is explored from stellar evolution 
models as well as gyrochronologically. As the photodynamical 
modelling yields precise densities, our derived values are also 
the subject of discussion. Furthermore, the available radial velocity 
measurements of this system have not been mentioned in this paper; 
the reasons behind this choice will be addressed here as well. The 
last point of this section deals with the innermost confirmed planet 
of the system, that is not included in the analysis. Finally, we discuss 
the possibility of detect other planets in the system by means of the 
observed TTVs of Kepler-9b/c.

\subsection{Dynamical stability}
A dynamical analysis leads naturally to the question of the long term 
stability of the derived planetary system, as an unstable result should 
not be considered as a viable model, contradicting
the long lifetime of the system. To test the stability of our 
results for the Kepler-9 system, our best photodynamical solution was 
extended in time up to $1\;$Gyr.
For this purpose we used the second-order mixed-variable symplectic 
algorithm implemented in the \texttt{Mercury6} package by 
\citet{Chambers1999}. This is the same integrator used in our 
photodynamical model, also including the post-Newtonian correction 
\citep{1995PhRvD..52..821K}. The time step size we used was $0.9\;$days, 
which is slightly smaller than a twentieth of the period of the 
innermost planet considered in our dynamical analysis, Kepler-9b. 
This step size gives a good compromise between reasonable computation 
time and small integration errors. We find that over the integration time
the modelled planetary system remains stable. Given the architecture of 
the system, this was expected, and we can assume that
the very similar good results from MCMC modelling should remain stable as well.

\subsection{Transit timings}
After the {\it Kepler} observations, the variation of good MCMC
models in comparison to the best data set III model show different 
amplitudes as time progresses (see for instance the time range around 
2014-2015, and around 2018-2020, Fig.~\ref{fig:ttv}). These variations
are illustrated by the grey areas in different shades for the 
modelling of the different data sets, from light to dark grey
corresponding to data sets I, II, and III.
At the specific times previously mentioned, the variations increase 
for both planets. This behaviour appears when the O-C has a positive 
slope for Kepler-9b, and a negative slope for Kepler-9c. At these 
places the gradient of the TTVs is larger in comparison 
to the parts where Kepler-9b shows decreasing TTVs and when Kepler-9c 
shows increasing TTVs. A larger gradient leads to a larger 
uncertainty in the predictions. Despite the lower precision in 
comparison to the space-based {\it Kepler} data, the new ground-based 
KOINet observations helped to set tighter constrains on the 
modulation of the timings. 
Unfortunately, besides one observation, we missed the chance 
to observe transits in the phase of higher variation amplitude 
in 2014. The next time with a higher amplitude starts in 2018, 
that means another few observations during 2018, especially 
Kepler-9c transits, will help to tighten the modulation even more.

\subsection{The disappearance of Kepler-9c transits}
\begin{figure}
\resizebox{\hsize}{!}{\includegraphics{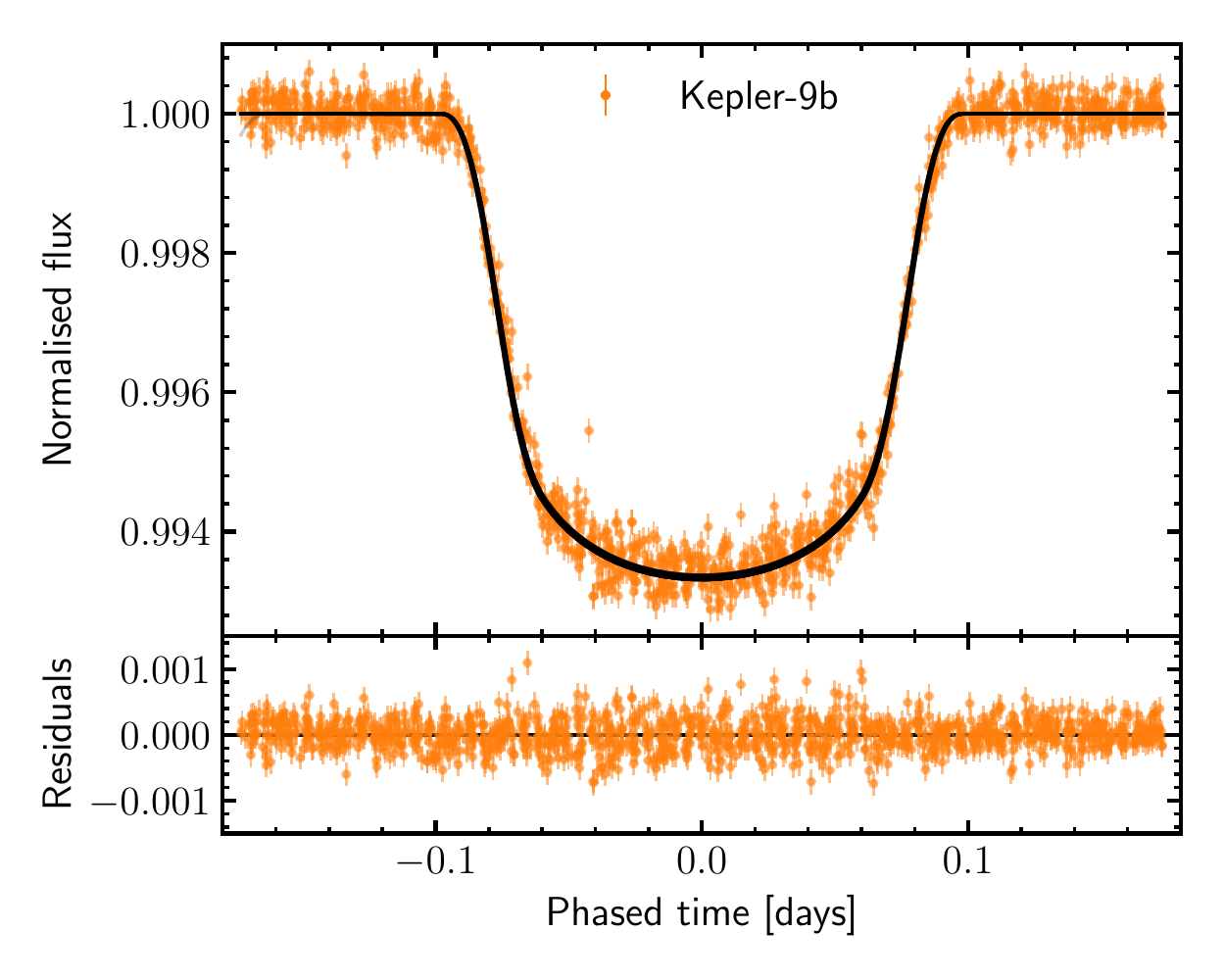}}
\resizebox{\hsize}{!}{\includegraphics{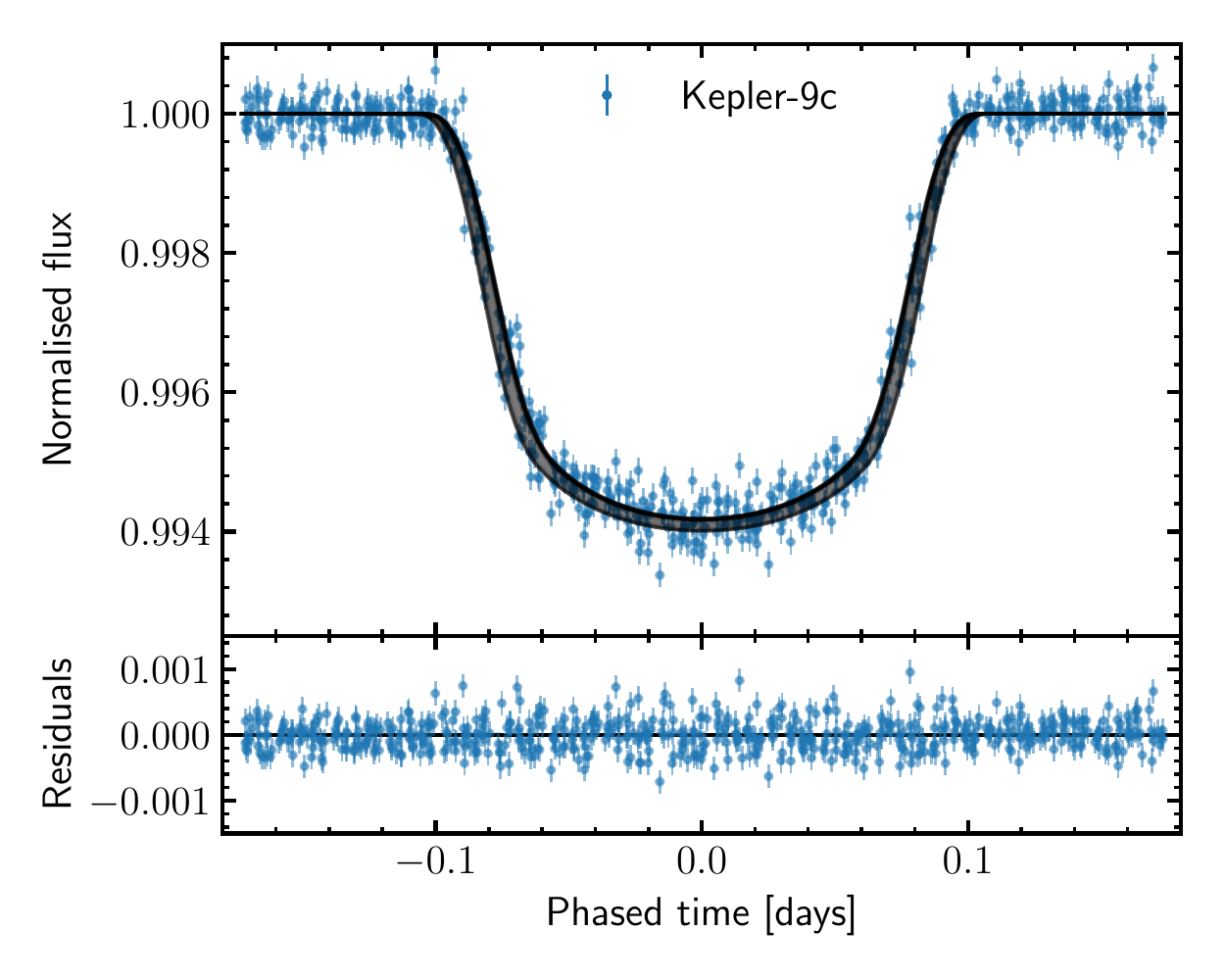}}
\caption{All long cadence {\it Kepler} (quarter $1-17$) data per 
planet aligned by the transit time in orange for Kepler-9b and in 
light blue for Kepler-9c. In grey is the best photodynamical model 
of the full data set, but calculated for only these {\it Kepler} long 
cadence data and aligned respectively. The bottom of each figure 
shows the residuals.}
\label{fig:alltransits}
\end{figure}

One of the advantages of our photodynamical modelling is the physical 
consistency in modelling variations in the transit shape due to 
variations in the transit parameters. These variations can be 
explained by the dynamical interaction of all 
objects in the system. Figure~\ref{fig:alltransits} shows these 
variations in the transit shape. Plotted are the transit light curve 
data per planet shifted by their individual transit time. For a 
better visualisation of this effect, we plotted only {\it Kepler} quarter 1 
through 17 long cadence data. The higher scatter of short cadence 
data would lead to a larger range in flux. In turn, the variation of the 
model would appear diminished due to the larger data range. The 
model variation is shown in black, which is the best model for each of the 
transits modelled in data set III, shifted to a common time of transit. 
For Kepler-9c especially there is a clear variation in the transit 
shape visible, both in transit depth and transit duration.

The variations in transit shape is not only best visible on 
long cadence data, but also most significant. The same TTV model 
with an averaged transit shape model gives a 8\% worse reduced $\chi^2$ on 
data set I. On data set II and III the difference in the reduced $\chi^2$ is 
only of the order of 0.5\%. Nevertheless, photodynamical modelling has the advantage 
of consistently model the TTVs with the transit shape determining 
parameters, that is mainly the inclination. 

\begin{figure*}
\includegraphics[width=9cm]{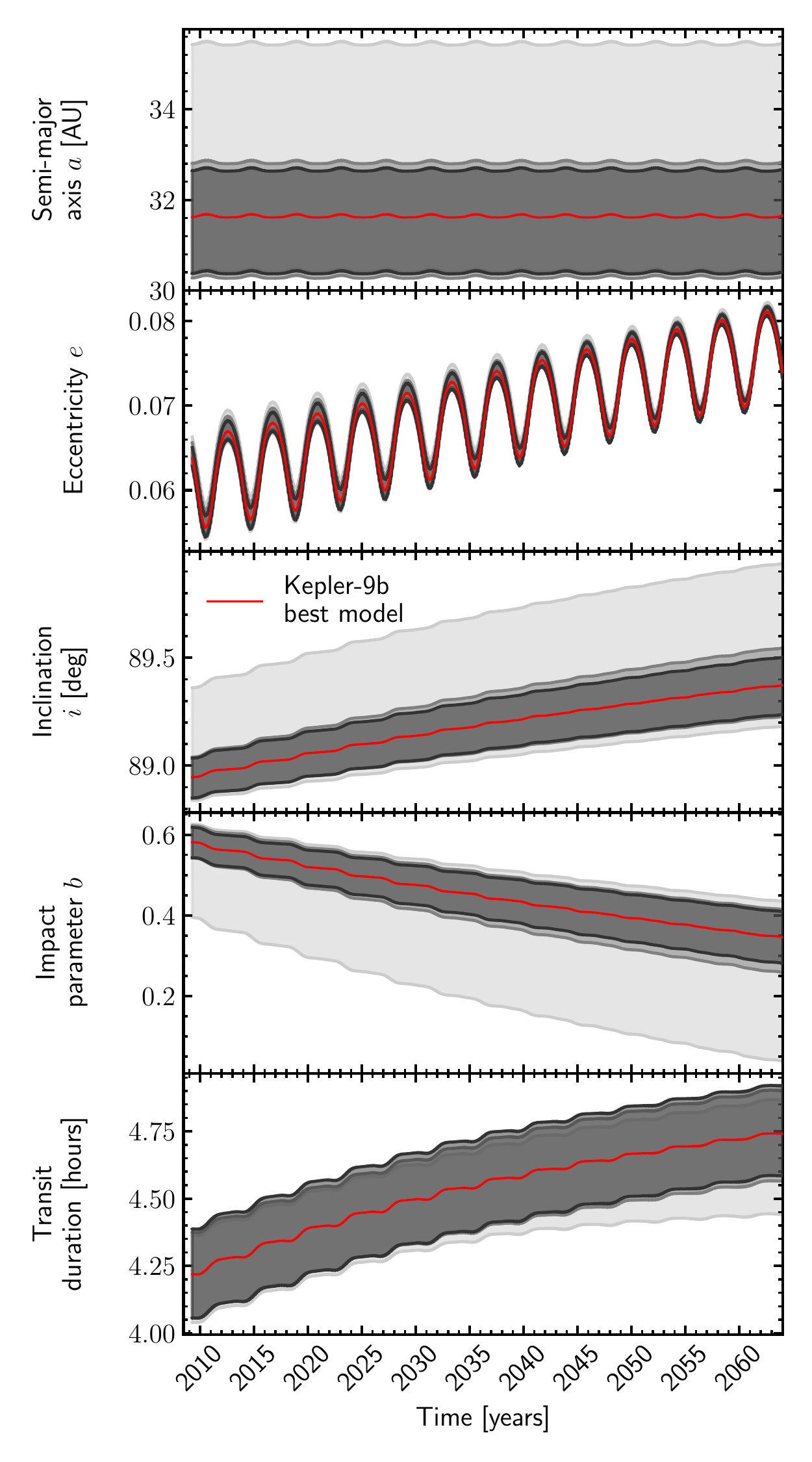}
\includegraphics[width=9cm]{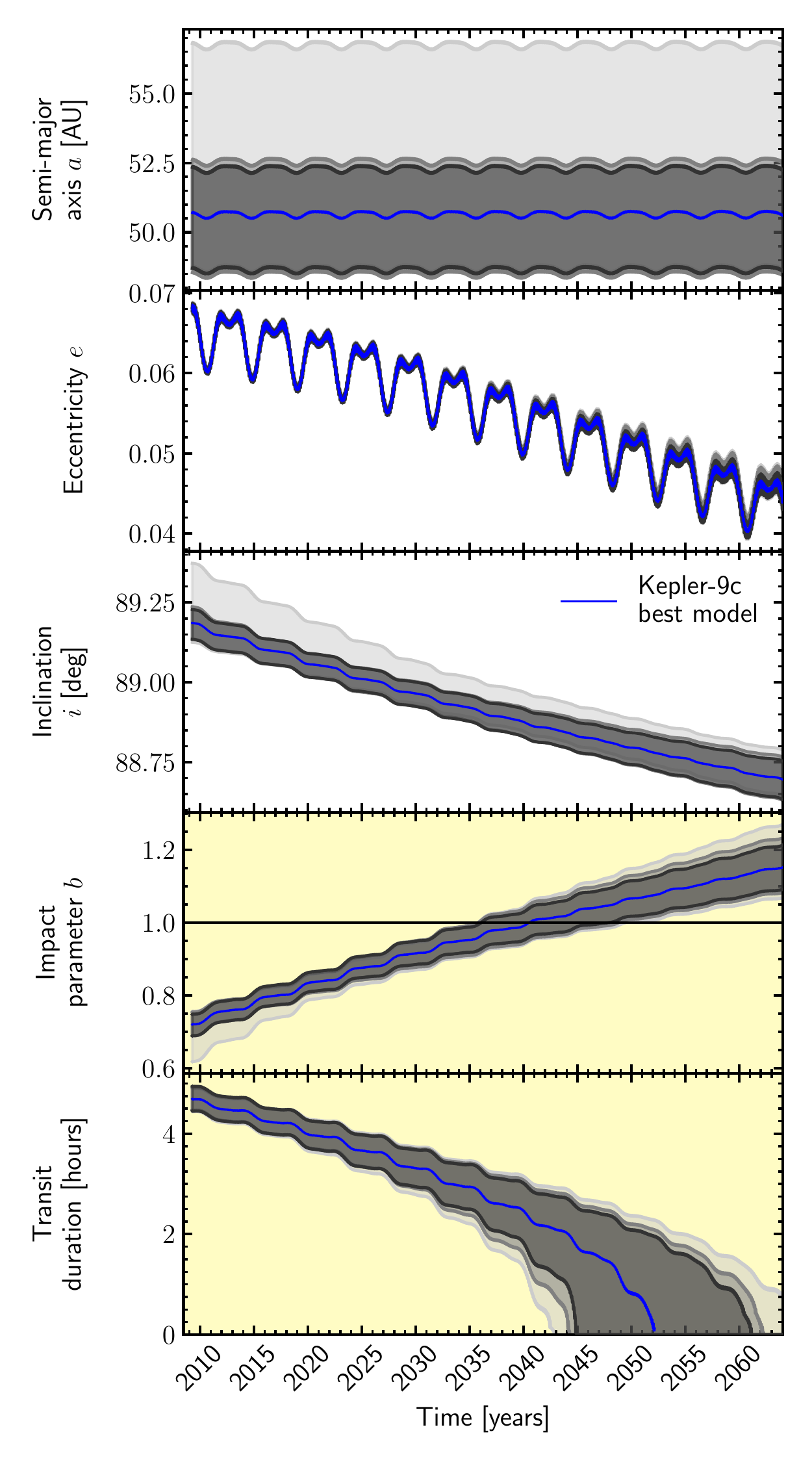}
\caption{Top: the extrapolation of the planets semi-major axis 
until 2065 for the best model in red (Kepler-9b) and blue 
(Kepler-9c) and 
in grey areas the 99.74\% confidence interval of 1000 randomly 
chosen good models for the different data sets. Light grey corresponds 
to the modelling of data set I, middle grey to data set II and 
dark grey to data set III respectively.
Second: the extrapolation of the planets eccentricity. 
Third: the extrapolation of the planets inclination.
Fourth: the extrapolation of the calculated impact parameter.
Fifth: the extrapolation of the calculated transit duration.
The background of the impact parameter and the transit duration 
is coloured for highlighting the place where the prediction of 
disappearing transits comes from.}
\label{fig:parameters}
\end{figure*}
The observed variation in the transit shape of Kepler-9c 
leads us to examine the evolution of transit parameters 
over time. Figure~\ref{fig:parameters} shows the variations of the 
semi-major axis, the eccentricity and the inclination with the 
predictions for the next 50 years. The predictions for the 
inclination of Kepler-9c show a continuous decrease, so that 
both the derived impact parameter $b$ and the transit duration indicate the 
disappearance of the transits around the year 2052. This behaviour is 
shown in Fig.~\ref{fig:parameters} as well. A long term inspection 
reveals the variations in inclination to be a periodic effect, so 
that the transits will return around 2230 again (see 
Fig.~\ref{fig:parameters_long} in the appendix).
Through the decreasing inclination, within the next 35 years we 
will have the opportunity to map the high latitudes and hence measure 
the limb of Kepler-9 with frequent transit observations of planet~c. 
In these higher latitudes the transit spends more time at the limb than
in the case of a passage of the mid-point of the star. This fact could help in getting 
more information of the atmospheric structure at the limb.

On the other 
hand, for planet b an increasing inclination in the next 100 years is 
predicted (see Fig.~\ref{fig:parameters_long} in the appendix). 
\citet{Wang2017b} measured a stellar spin alignment with the planetary 
orbital plane on a high probability. As this result might be affected by 
stellar spot crossings, more Rossiter–McLaughlin measurements are necessary 
for a confirmation \citep{Oshagh2016}. Nevertheless, assuming spin 
alignment, Kepler-9b will scan the latitudes between above $30^\degree$ now 
down to around $10^\degree$. With Kepler-9 being a solar analog, a spot 
appearance similar to the sun between $0^\degree$ and $30^\degree$ is a 
reasonable assumption. Under those circumstances, measuring spot crossings 
by Kepler-9b in precise transit observations in the future has a high 
probability. Such detections would lead to a starspot distribution 
measurement like in the work of \citet{Morris2017} in which the system 
analysed,
HAT-P-11, is known to be highly misaligned. Hence, an even more similar 
analysis is possible if the spin alignment of Kepler-9 is not confirmed,
in which case there could possibly be spot contamination in the existing 
transit observations. Spot crossings are not resolvably measured, so a higher 
accuracy would be necessary for this end. 

The existing coverage 
of latitudes by transit observations is illustrated in 
Fig.~\ref{fig:Latitudes} under the assumption of a stellar spin alignment 
with the orbital plane. The red lines refer to Kepler-9b and blue to Kepler-9c 
transits. The track is extracted from the best photodynamical model of 
data set III. The uncertainties in these tracks are retrievable from the 
impact parameter shown in the fourth row of Fig.~\ref{fig:parameters}. 
The yellow circular disk illustrates the star and the orange area the possible 
star spot occurrence ranges assuming a similar behaviour to the sun.

The precise {\it Kepler} data allow us to model the quadratic limb 
darkening of the star. As a result, from modelling data set III the
derived limb darkening coefficients are $c_1=0.35\pm0.05$ and 
$c_2=0.27\pm0.07$. Figure~\ref{fig:corner_radius} shows that these 
two coefficients are highly anti-correlated. This result is 
consistent with \citet{Mueller2013}, who investigated the quadratic 
limb darkening of {\it Kepler} targets. Additionally, the values suit the 
literature values given in the \texttt{NASA Exoplanet Archive} 
\citep{NASAexopA}. The results from modelling data set I demonstrate 
that using only long cadence {\it Kepler} data is not sufficient to model 
the quadratic limb darkening of Kepler-9. Nonetheless, the derived 
values of $c_1=0.28\pm0.05$ and $c_2=0.41\pm0.09$ fit the 
anti-correlation derived by modelling data set III. This 
anticorrelation is illustrated in the parameter correlation 
plot in Fig.~\ref{fig:corner_radius}. Consequently, the discrepant 
values lead to different results for the stellar radius and the 
planet-star radius ratios.

In order to check for model-dependent influences on the resulting 
evolution of the system parameters, we investigated the differences 
between Newtonian gravity and the inclusion of a post-Newtonian 
correction. An analysis was done for the influence on resulting 
photodynamical models for the Kepler-9b and c. Including the 
post-Newtonian correction decreased the parameter uncertainties in 
the second significant figure and the reduced $\chi^2$ in the fifth. 
The differences are too small to discriminate the models. The 
future predictions for change in inclination and the transit times 
behave very similarly.

\subsection{On the stellar radius, mass and age}
\label{subsec:stellarevol}
\begin{figure}
\resizebox{\hsize}{!}{\includegraphics{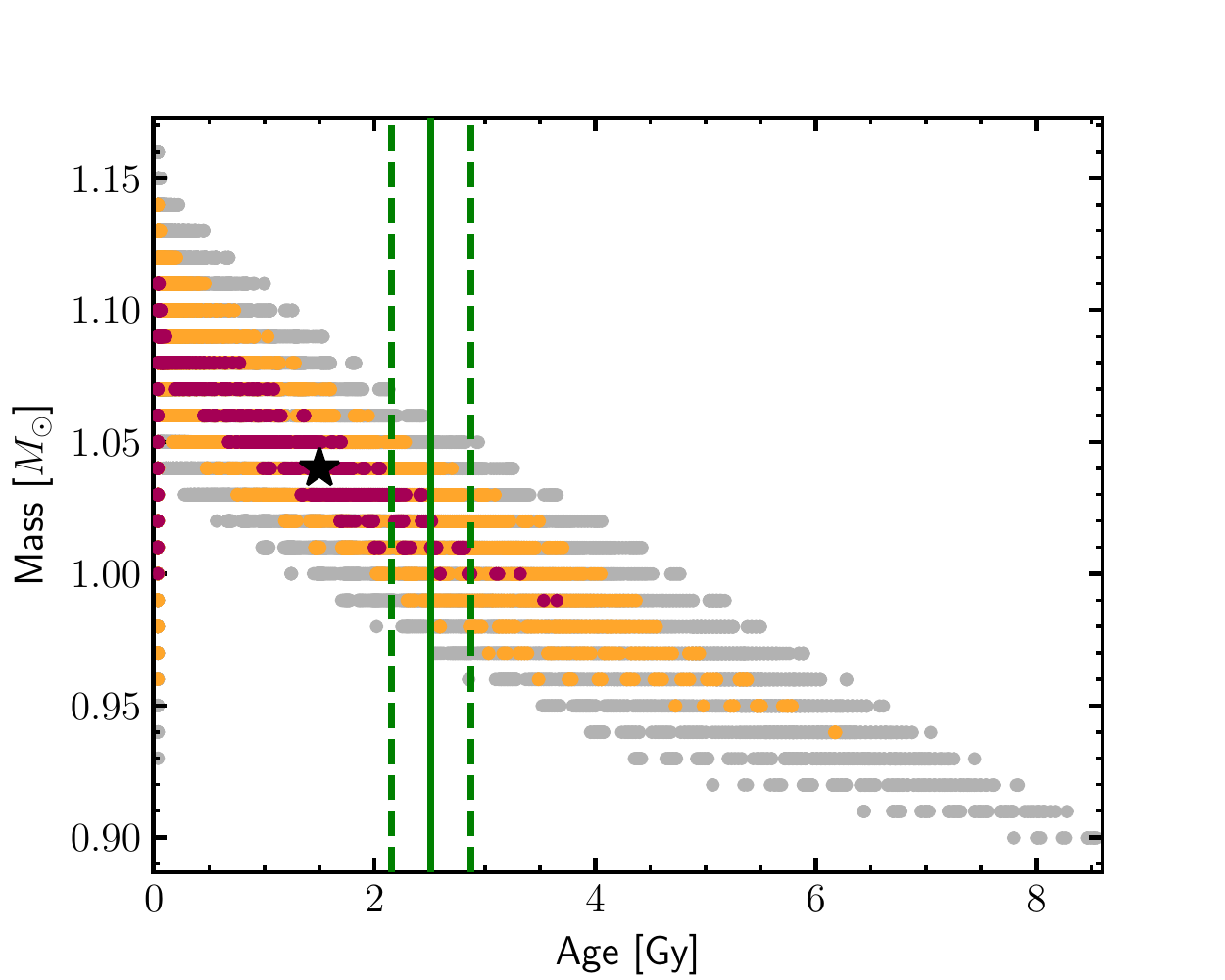}}
\caption{Mass-age diagram of Kepler-9 from MESA stellar evolution
models (MIST). The black star and the red, orange, and grey dots correspond 
to the best matching value and the $1\sigma$, $2\sigma$,
and $3\sigma$ areas derived from results on the density of the data set III
photodynamical modelling and from new literature values of the effective 
temperature, the surface gravity, and the metallicity by \citet{Petigura2017}.
The gyrochronological age is indicated by the green solid line and 
its 1-$\sigma$ range with the green dashed lines.}
\label{fig:MassAge2}
\end{figure}
\begin{figure}
\resizebox{\hsize}{!}{\includegraphics{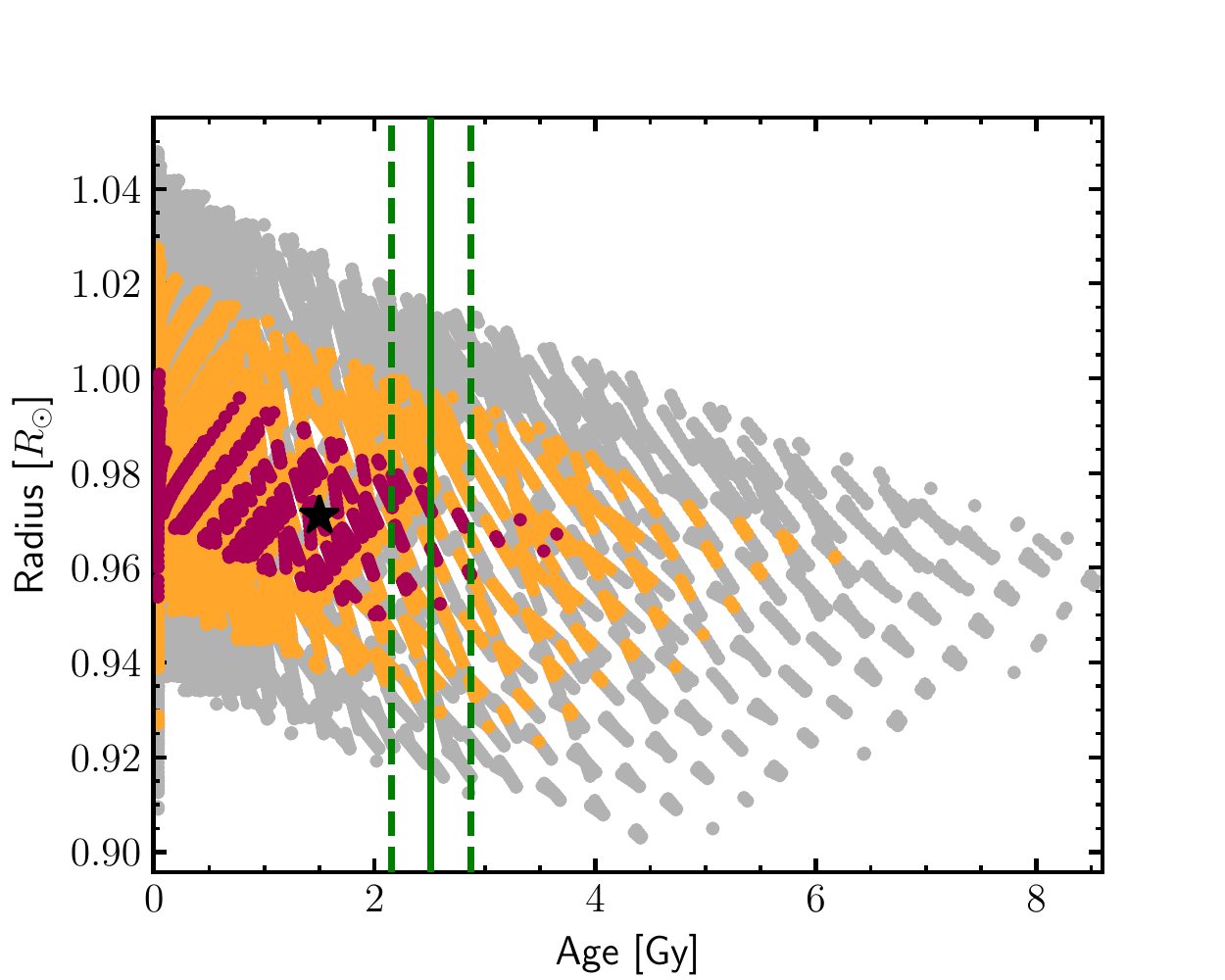}}
\caption{Radius-age diagram of Kepler-9 from MESA stellar evolution
models (MIST). The black star and the red, orange, and grey dots correspond 
to the best matching value and the $1\sigma$, $2\sigma$,
and $3\sigma$ areas derived from results on the density of the data set III
photodynamical modeling and from new literature values of the effective 
temperature, the surface gravity, and the metallicity by \citet{Petigura2017}.
The gyrochronological age is indicated by the green solid line and 
its 1-$\sigma$ range with the green dashed lines.}
\label{fig:RadiusAge2}
\end{figure}
Applying our photodynamical analysis to data set III we determined a 
stellar density of $\rho_S = 1.603\pm0.061\;\text{g cm}^{-3}$. As described
in section~\ref{sec:photdyn} we modelled the stellar radius instead of the 
density. However, the transit measurements constrain the stellar
density \citep{AgolFabrycky2017}. With a fixed stellar mass, the density 
can be determined straightforwardly.
Our modelled density is almost 50\% higher than the prior estimate 
($\rho_S = (0.79 \pm 0.19)\;\rho_\sun\;\hat{=}\;1.12\pm0.27\;\text{g cm}^{-3}$)
by \citet{Havel2011}. 
The authors derived this value from the TTV analysis 
of the first 3 quarters \textit{Kepler} observations by \citet{Holman2010}.
With this density, the stellar mass, radius and age was determined 
by stellar evolution models. Our considerably higher derived density
motivated a new, similar study of Kepler-9.

We used the stellar density and the known stellar parameters of an
effective temperature $T_\text{eff}=5777\pm61\;$K, surface gravity 
$\log g =4.49\pm0.09\;$ and metallicity 
$\text{Fe}/\text{H}=0.12\pm0.04$ \citep[which classifies Kepler-9 as solar analog, see][]{Holman2010, Havel2011} 
to determine the age, the mass and the radius of the star by stellar 
evolution models. The results are presented in the appendix (similar to the 
below derived Fig.~\ref{fig:MassAge2} and Fig.~\ref{fig:RadiusAge2}) in a 
mass-age diagram in Fig.~\ref{fig:MassAge} and in a radius-age diagram 
in Fig.~\ref{fig:RadiusAge}. We extracted the corrsponding values from 
the interpolated MESA \citep{Paxton2011, Paxton2013, Paxton2015} 
evolutionary tracks by MIST \citep{Dotter2016, Choi2016}. 
We derive a stellar mass of $m_S=1.06_{-0.05}^{+0.06}\;M_\sun$, 
a radius of $R_S=0.977_{-0.024}^{+0.031}\;R_\sun$ and an age of 
$\tau_\text{Evol}=0.95_{-0.92}^{+1.88}\;$Gyr. 

Recent HIRES observations by \citet{Petigura2017} of more than 
1000~KOIs led to the correction of the Kepler-9 stellar parameters 
to $T_\text{eff}=5787\pm60\;$K, $\log g =4.473\pm0.1\;$, and  
$\text{Fe}/\text{H}=0.082\pm0.04$. Although very similar, the lower 
metallicity leads to slightly different results. With these new values 
we determined the stellar mass to $m_S=1.04_{-0.05}^{+0.07}\;M_\sun$, 
the radius to $R_S=0.971_{-0.021}^{+0.030}\;R_\sun$ and the stellar 
age to $\tau_\text{Evol}=1.49_{-1.47}^{+2.15}\;$Gyr. The corresponding
diagrams can be found for mass vs.\ age in Fig.~\ref{fig:MassAge2} 
and for radius vs.\ age in Fig.~\ref{fig:RadiusAge2}. We like to note that 
mass and radius for both parameter sets are in agreement within $1\,\sigma$. 
The derived age of 1.5\ Gyr, however, is in better agreement with the 
gyrochronological age derived below. With these new values
for the stellar mass and radius we corrected the modelled planetary 
masses, semi-major axes, and radii, which can be found in the sixth 
column of Table~\ref{table:results}.

More recently, the second Gaia data release (Gaia DR2) was carried out 
\citep{Gaia2016,Gaia2018}. The there derived effective temperature of
$T_\text{eff}=5750_{-130}^{+250}\;$K fits the HIRES value within the $1\sigma$ 
range, so does the stellar radius with $R_S=0.977_{-0.080}^{+0.045}$. 
The values have comparatively higher uncertainties. The distance of 
Kepler-9 is determined to $p=1.563\pm0.017\;$mas by Gaia DR2.

To test the results of the stellar evolution model analysis, 
we determined the gyrochronological age of 
Kepler-9. For this, we computed a periodogram of Kepler-9's full long 
cadence photometry \citep{Lomb,Scargle,LombScargle}. The highest 
power peak corresponds to 16.83 $\pm$ 0.08 days. The period and 
error correspond to the mean and standard deviation obtained fitting 
a Gaussian to the highest periodogram peak. On Kepler-9, typical 
photometric variability due to spot rotation has an amplitude of 
5~ppt, well above the photometric noise.

To determine Kepler-9's age we made use of 
\citet{Barnes2007,Barnes2009}'s gyrochronological estimate:
\begin{equation}
    \log(\tau_\text{Gyro}) = \frac{1}{n}[\log P - \log a - b \times \log (\text{B-V} - c)]\, ,
\end{equation}
\noindent for $a = 0.770 \pm 0.014$, $b = 0.553 \pm 0.052$, 
$c = 0.472 \pm 0.027$, and $n = 0.519 \pm 0.007$. Assuming 
B-V~$= 0.642$, and following \citet{Barnes2009} error estimates, the 
derived gyrochronological age for Kepler-9 is 2.51~$\pm$~0.36~Gyr. 
This age is indicated in the mass-age and radius age diagrams 
(Fig.~\ref{fig:MassAge2}, Fig.~\ref{fig:RadiusAge2}, Fig.~\ref{fig:MassAge}, 
Fig.~\ref{fig:RadiusAge}) by green lines, solid for the the median value and 
dashed for the 1-$\sigma$ range. The gyrochronological age is slightly higher
than the age indicated by stellar evolution models,
but the values agree within the 1-$\sigma$ range.

\subsection{The stellar and planetary densities}
\begin{figure}
\resizebox{\hsize}{!}{\includegraphics{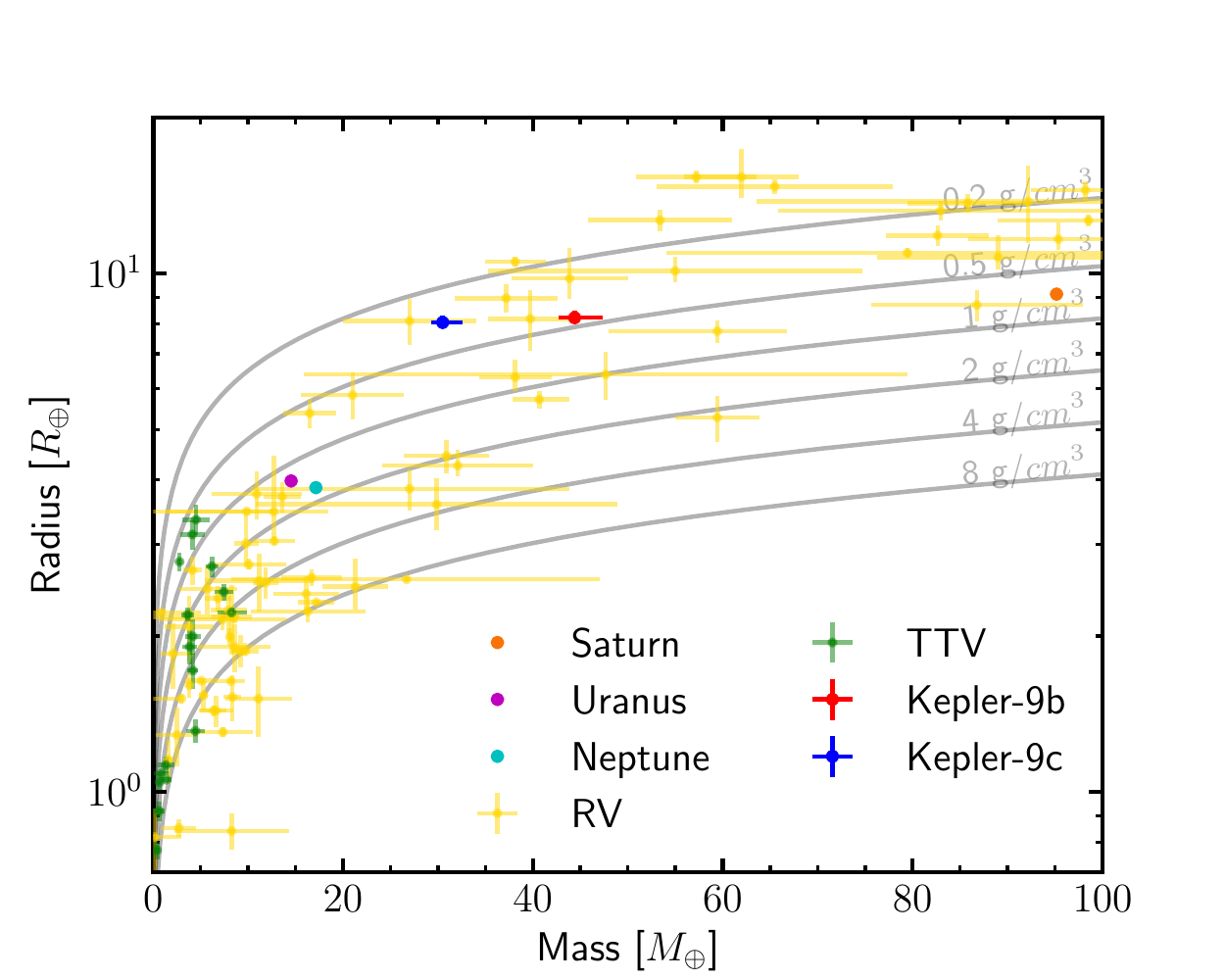}}
\caption{Mass-Radius diagram for known planets with masses up to 
$100\;M_\Earth$. In yellow are the planets with mass measurements 
obtained by radial velocities and in green the planets with mass 
measurements obtained from TTVs. The data are given by the 
\texttt{The Extrasolar Planets Encyclopaedia}. Our results are shown 
in red (Kepler-9b) and in blue (Kepler-9c). For comparison also the 
values of Saturn, and Uranus and Neptune are shown, the Neptune-like 
planet pair of our solar system.}
\label{fig:MassRadius}
\end{figure}
In addition to the stellar density, the photodynamical analysis provides 
strong constraints on the planetary densities. As a result of the 
analysis performed on data set III we obtained densities of 
$\rho_b =0.439 \pm 0.023 \;\text{g/cm}^3$ for Kepler-9b and 
$\rho_c= 0.322 \pm 0.017 \;\text{g/cm}^3$ for Kepler-9c. In 
Fig.~\ref{fig:MassRadius} our results are compared to literature 
values from \texttt{The Extrasolar Planets 
Encyclopaedia}\footnote{\url{http://www.exoplanet.eu/}} for planets 
with similar properties. Color-coded are the mass measurements 
obtained from radial velocities in yellow, and from TTVs in green. In 
this regime, that is the regime of Neptune-like planets, the density 
measurements of Kepler-9b/c are to date the most precise ones outside
the solar system.

To rule out biased results for stellar radius and planet-star radii 
ratios caused by the photometric variability of Kepler-9, we checked for 
variability in the residuals of the transit light curves. 
To be consistent we chose the high-precision, well-sampled
\textit{Kepler} short cadence data for this analysis. The scatter 
of the residuals inside the transit is slightly larger than outside 
the transit. The standard deviation inside the transit is 
$\text{std}_\text{inside}=0.001049$, while outside is 
$\text{std}_\text{outside}= 0.001027$ for the best model of 
data set III. That means 2\% difference between inside and outside
transit. We did not find any periodicity inside the transit residuals, 
potentially caused by star spots. Equivalently, the transit time residuals 
do not show a periodic variability. Nevertheless, the higher scatter inside
transit possibly results from unresolved stellar spot crossings. The 
planet-star radii ratio determination is affected within its uncertainties.
With the absence of measurable star spots and the small differences in 
standard deviation between inside transit and outside, a systematic error 
in the radius determination seems to be negligible. The planetary densities 
are therefore also well determined.

Figure~\ref{fig:MassRadius} shows the similarity of the Kepler-9b/c 
planets to Saturn in radius. The masses are less than half the value 
of Saturn, resulting in smaller densities. Their low density implies Kepler-9b/c
should be classified as hydrogen-helium gas giants. The formation of the planets
happened most likely in the outer region of the system. Through 
converging migration the planets could be brought in 
the near 2:1 mean motion resonance in close proximity to the host star
\citep[e.g. shown by][]{Henrad1982,Borderies1984,Lemaitre1984}.
It has been shown that such formation scenarios can result in stable
resonant orbits with the outer planet having only about half the mass 
of the inner one \citep{DeckBatygin2015}.

\subsection{The radial velocity measurements}
\begin{figure}
\includegraphics[width=8.5cm]{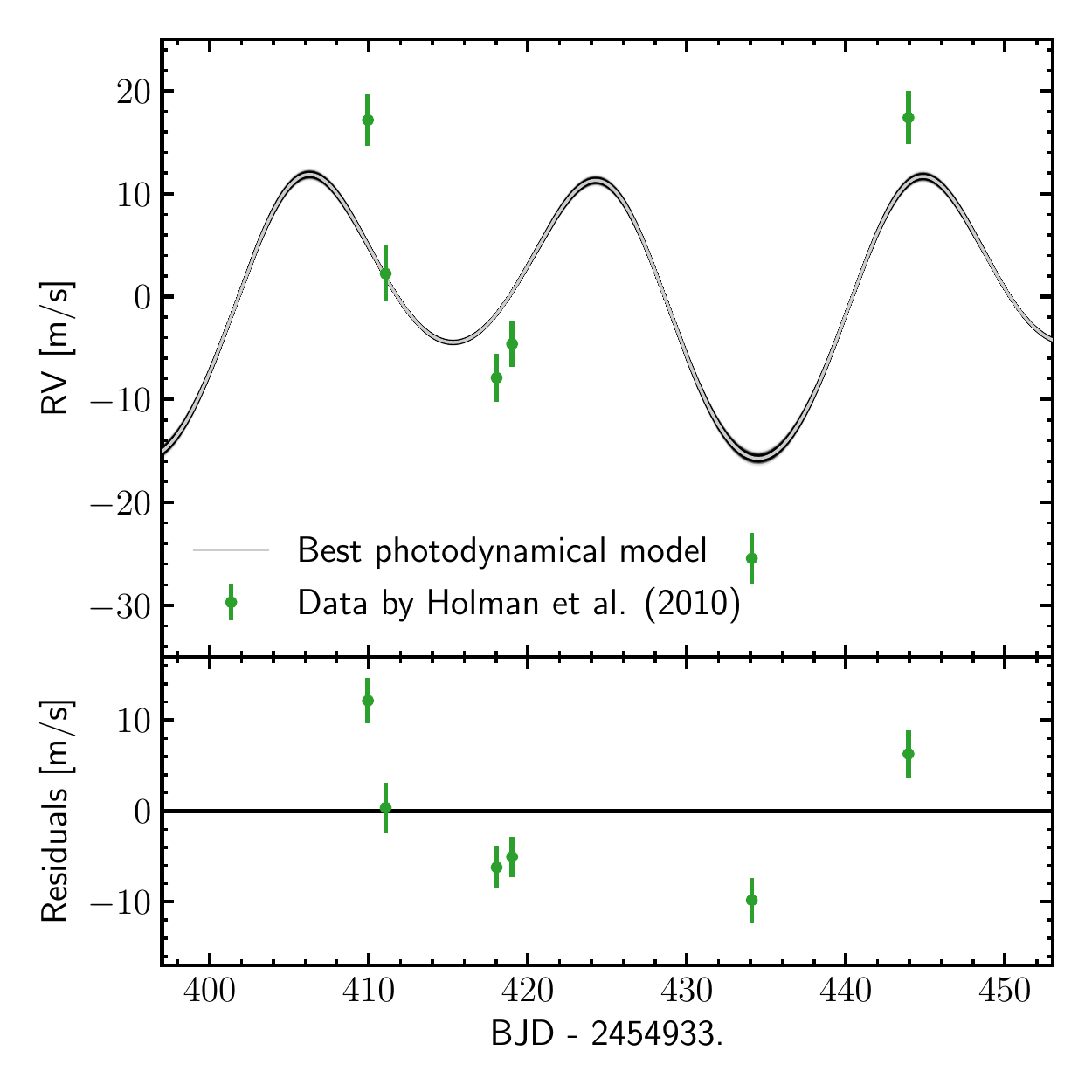}
\caption{Results from photodynamical modelling of data set III
on the radial velocity measurements by \citet{Holman2010}.}
\label{fig:RV}
\end{figure}

In our analysis we did not considered the radial velocity (RV) 
measurements by \citet{Holman2010} for Kepler-9. The reasons are the 
small number of measurements, the short time span of the observations, 
as well as the large discrepancy between a dynamical model to the 
TTVs and the RV data. Nevertheless, from our photodynamical model we 
calculated an RV model. Simulated RV models from the results of 
modelling the full transit dataset are shown together with the data 
in Fig.~\ref{fig:RV}. The best model has a $\chi^2=56.94$ for 
the 6 RV measurements. As pointed out by \citet{DreizlerOfir2014}, we 
also see a similar discrepancy between the dynamical model and these 
measurements.  Additional evidence in favor of the TTV model comes
from the short timescale chopping variations seen in planet 9b due
to 9c:  the amplitude of chopping prefers a smaller mass for planet
9c \citep{Deck2015}, which, of course, is included in the full
photodynamical constraint.

The most evident reason for this discrepancy is the activity of 
Kepler-9. A jitter factor would be necessary for including these data 
in the analysis. A detailed analysis of the activity of the star and 
the integration of the RV measurements is however beyond the scope of 
this paper. 

In addition, the recently obtained RV measurements listed in the 
HARPS-N archive\footnote{\url{http://archives.ia2.inaf.it/tng/faces/search.xhtml?dswid=9814}}, 
but marked as proprietary could help to 
constrain the RV behaviour of Kepler-9 better. 
Figure~\ref{fig:RV_pred} shows a prediction of the Kepler-9 radial 
velocity based on our model constraints for the approximate time 
span of the new HARPS-N observations.

\subsection{An additional planet?}

To complete our analysis we also tested the influence of the third 
known planet, Kepler-9d \citep[confirmed by][]{Torres2011}, 
with a period $P_d=1.592960(2)\;$d on the 
dynamics of the system. We agree with \citet{DreizlerOfir2014} that 
it does not interact measurably with the two modelled planets, in the 
plausible mass regime \citep[$m_d=$~1-7~$M_\Earth$, ][]{Holman2010}. 
For a mass of $m_d=7\;M_\Earth$ the reduced $\chi^2$ does not improve 
and also the variations of the residuals of the transit times are of 
the same order. The amplitude in radial velocity measurements is of 
the order of $1\;{\rm m\,s}^{-1}$, far below the precision of the 
previous observations and currently infeasible for such a faint star 
as Kepler-9.

Adding another outer planet in a Laplace-resonance (4:2:1) to explain the 
deviations in the radial velocity measurements would require a rather 
high mass for the additional planet. Such a planet would have a far too large influence in the 
system's dynamics and is ruled out by the photodynamical analysis.
The fact that only 6 RV measurements are published makes it impossible 
to set constrains on further possible planets. Additional planets could
exist outside the Laplace-resonance that can explain the discrepancies
between transit and radial velocity measurements, and yet not influence 
the short-term dynamics much. 
In addition there is no periodicity in the transit 
timing residuals found, that would indicate an outer planet.

\section{Conclusions}
\label{sec:conclusion}
With this work we substantiated the importance of the KOINet. 
With its anticorrelated, large amplitude TTVs the Kepler-9b/c system
was chosen as benchmark system for the photodynamical modelling.
Although the dynamical cycle was almost covered by \textit{Kepler} 
observations, the 13 new transit observations 
led to better constrains on the composition of the system.
Concurrently, the capability of KOINet to complete a transit 
observation with long duration by using several telescopes around
the globe was confirmed. 
This is rounded out by the results of the photodynamical modelling.
The application to Kepler-9 revealed that the transits of the outer
planet will disappear in about 30 years. Furthermore, this dynamical
analysis of the combined photometric data, consisting of
{\it Kepler} long and short cadence data, complemented by ground-based
follow-up observation, led to the most precise planetary
density measurements of planets in the Neptune-mass regime.

From the decreasing inclination of Kepler-9c and increasing 
inclination of Kepler-9b we have the opportunity to map the different 
latitudes of the star. Hence, measurements of 
the limb and possibly star spots of Kepler-9 could be made possible
by precise, frequent transit 
observations within the next 35 years for the limb and 100 years for the 
star spots. 
Interspersed with frequent ground-based follow-up, transit
measurements from space that provide a high photometric 
precision would complement the stellar analysis. The promising 
predictions of this work make Kepler-9 an interesting target
for space missions like TESS \citep{TESS}, PLATO 2.0 \citep{PLATO} or 
CHEOPS \citep{CHEOPS}, though it is a rather 
faint object with Kepler magnitude of $K_p=13.803$. 

\begin{acknowledgements}
We acknowledge funding from the German Research Foundation (DFG) 
through grant DR 281/30-1. This work made use of PyAstronomy. CvE 
acknowledges funding for the Stellar Astrophysics Centre, provided 
by The Danish National Research Foundation (Grant DNRF106). Based 
on observations made with the Nordic Optical Telescope, operated 
by the Nordic Optical Telescope Scientific Association at the 
Observatorio del Roque de los Muchachos, La Palma, Spain, of the 
Instituto de Astrofisica de Canarias. We acknowledge support from 
the Research Council of Norway’s grant 188910 to finance service 
observing at the NOT. SW acknowledges support for International 
Team 265 (``Magnetic Activity of M-type Dwarf Stars and the Influence 
on Habitable Extra-solar Planets) funded by the International Space 
Science Institute (ISSI) in Bern, Switzerland. EA acknowledges
support from United States National Science Foundation grant 1615315. 
EH and IR acknowledge support by the Spanish Ministry of Economy and 
Competitiveness (MINECO) and the Fondo Europeo de Desarrollo Regional 
(FEDER) through grant ESP2016-80435-C2-1-R, as well as the support of 
the Generalitat de Catalunya/CERCA programme. The Joan Or\'{o} Telescope 
(TJO) of the Montsec Astronomical Observatory (OAdM) is owned by the 
Generalitat de Catalunya and operated by the Institute for Space 
Studies of Catalonia (IEEC). Based on observations 
collected at the Centro Astron\'omico Hispano Alem\'an (CAHA) at 
Calar Alto, operated jointly by the Max-Planck Institut f\"ur Astronomie 
and the Instituto de Astrof\'isica de Andaluc\'ia (CSIC). Based on 
observations obtained with the Apache Point Observatory 3.5-meter telescope, 
which is owned and operated by the Astrophysical Research Consortium. 
The Liverpool Telescope is operated on the island of La Palma by Liverpool 
John Moores University in the Spanish Observatorio del Roque de los 
Muchachos of the Instituto de Astrofisica de Canarias with financial 
support from the UK Science and Technology Facilities Council. 

\end{acknowledgements}

\bibliography{KOINet2}
\bibliographystyle{aa.bst}

\begin{appendix} 
\section{Additional plots and a table of the transit time predictions}

\begin{figure*}
\resizebox{\hsize}{!}{\includegraphics[width=.95\textwidth]{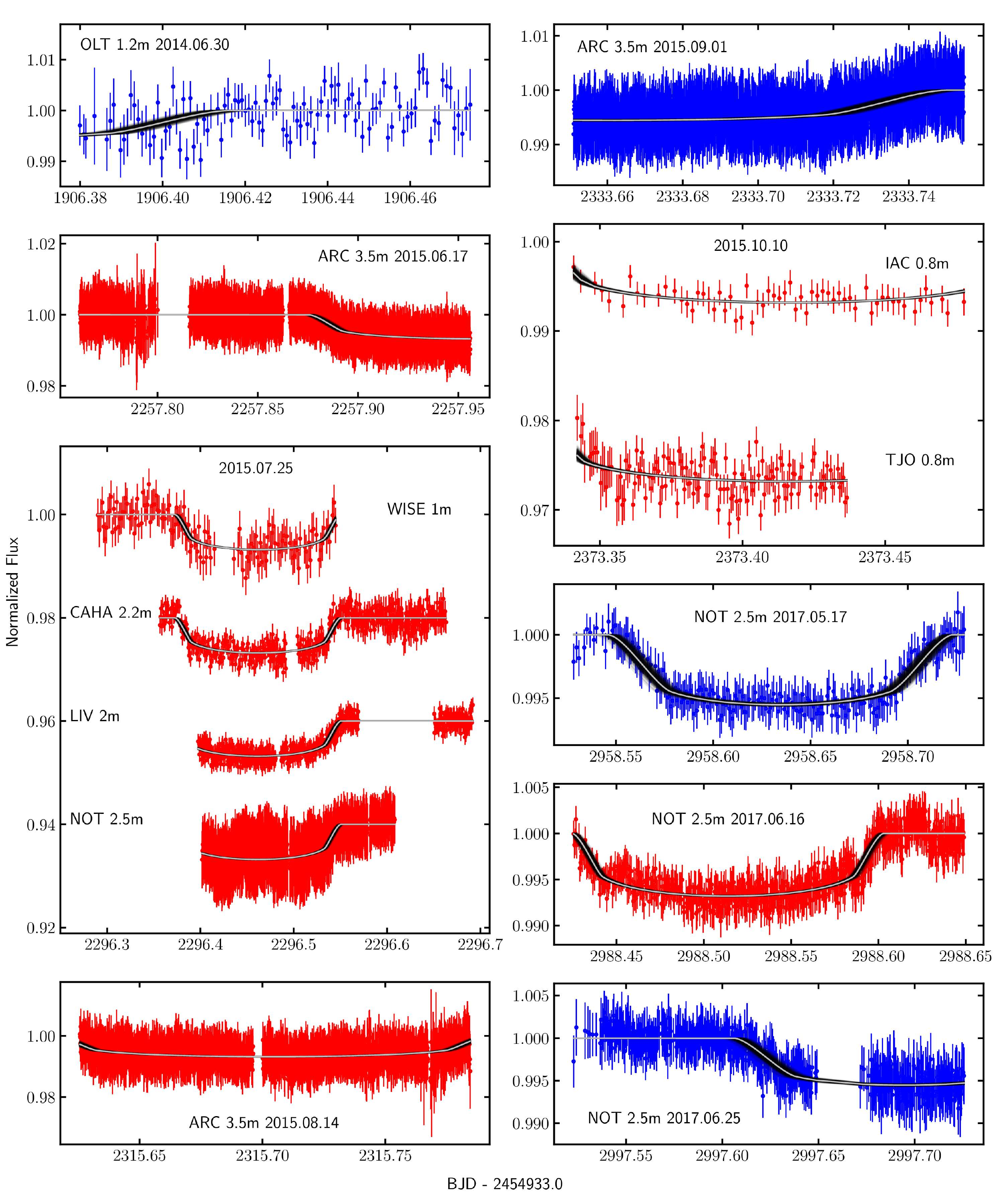}}
\caption{Our observed transits with the best data set III
model in grey and its variations by 500 randomly chosen good models
in black. Transit data in red belong to Kepler-9b and blue transit data  
correspond to Kepler-9c. The transits of dates with more than one observation
are artificially shifted for better visualiation and the belonging telescope 
is indicated.}
\label{fig:allnewtransits}
\end{figure*}

\begin{figure*}
\includegraphics[width=9cm]{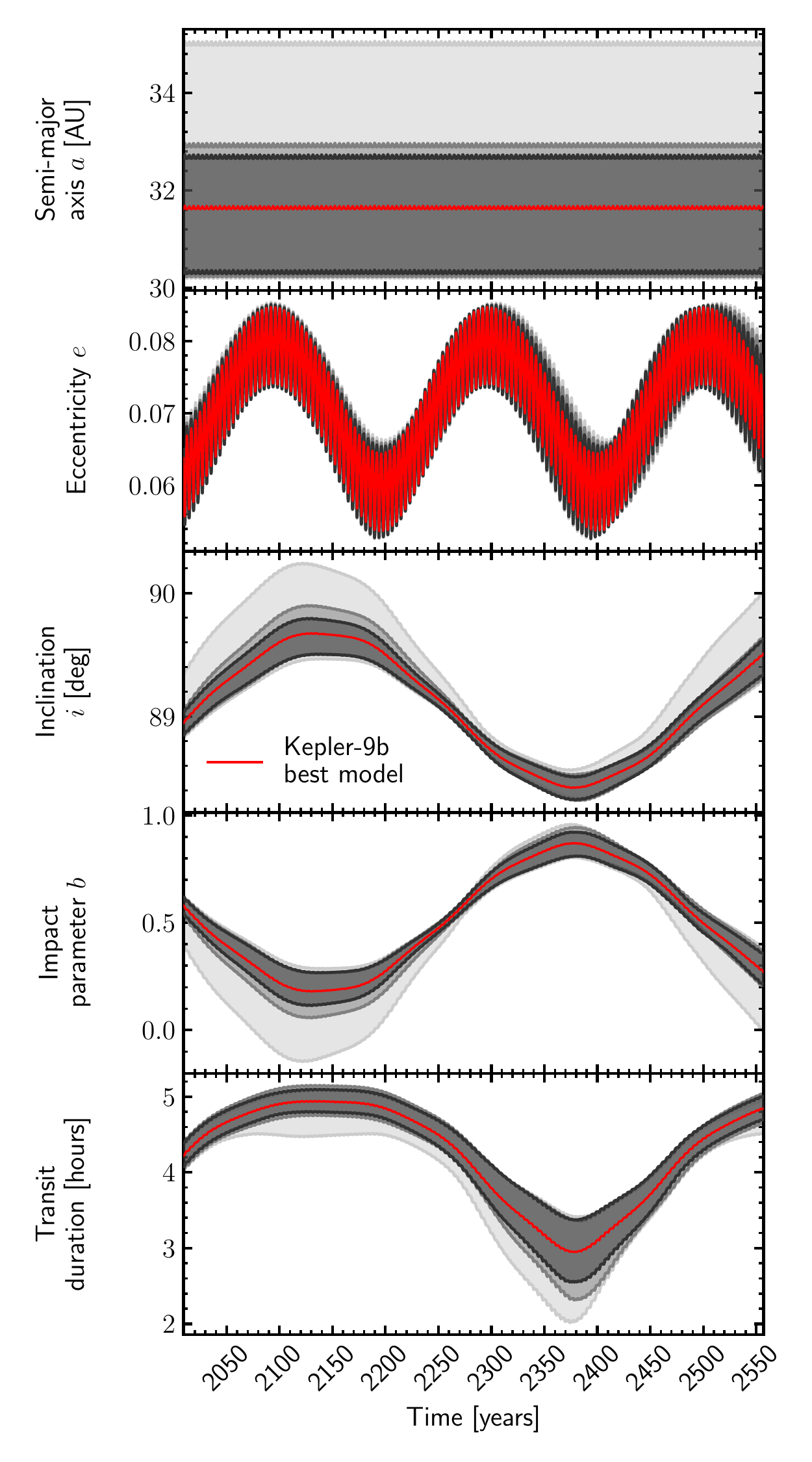}
\includegraphics[width=9cm]{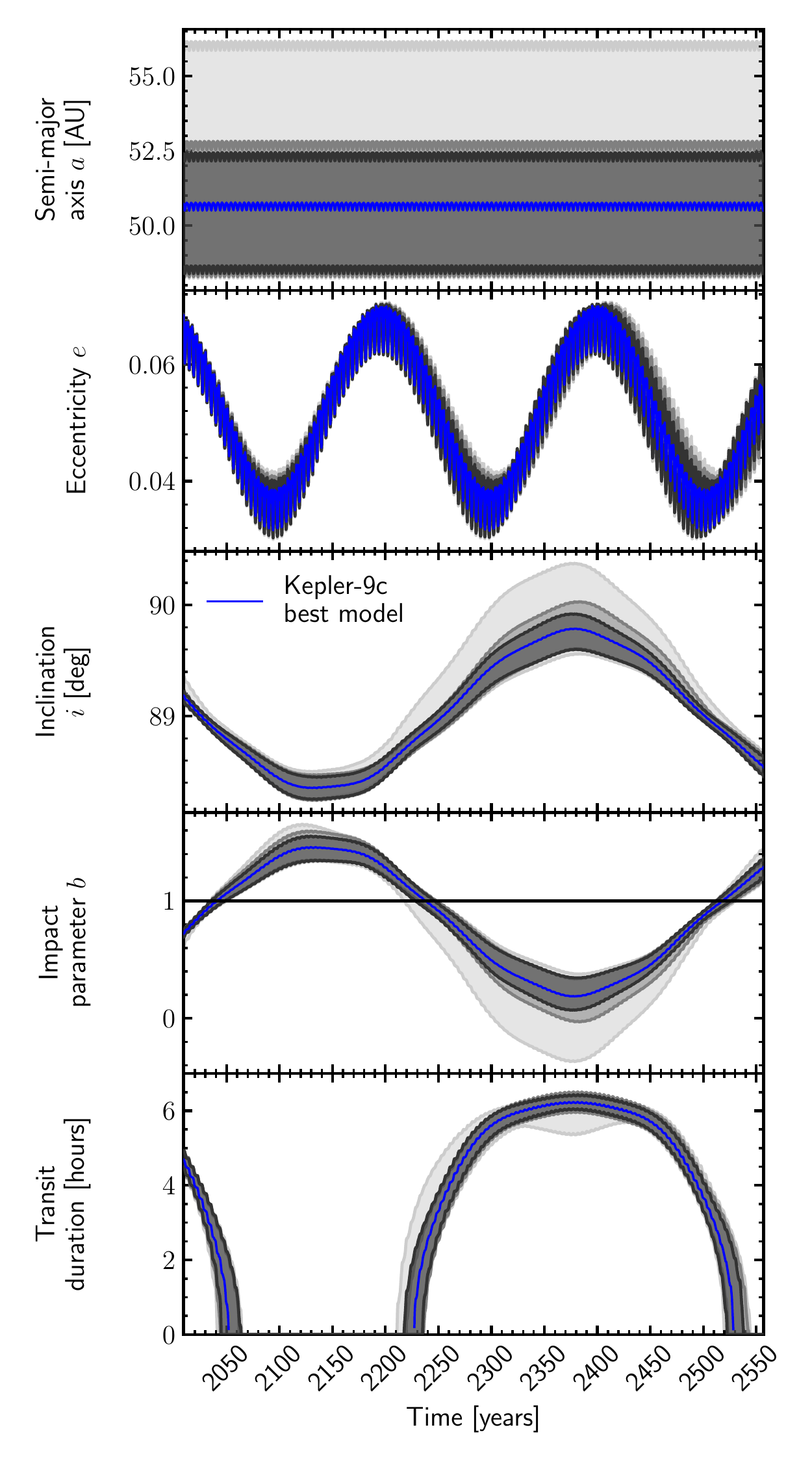}
\caption{Top: the extrapolation of the planets semi-major axis 
until 2550 for the best model in red (Kepler-9b) and blue 
(Kepler-9c) and 
in grey areas the 99.74\% confidence interval of 1000 randomly 
chosen good models for the different data sets. Light grey corresponds 
to the modelling of data set I, middle grey to data set II and 
dark grey to data set III repectively.
Second: the extrapolation of the planets eccentricity. 
Third: the extrapolation of the planets inclination.
Fourth: the extrapolation of the calculated impact parameter.
Fifth: the extrapolation of the calculated transit duration.}
\label{fig:parameters_long}
\end{figure*}

\begin{figure}
\resizebox{\hsize}{!}{\includegraphics{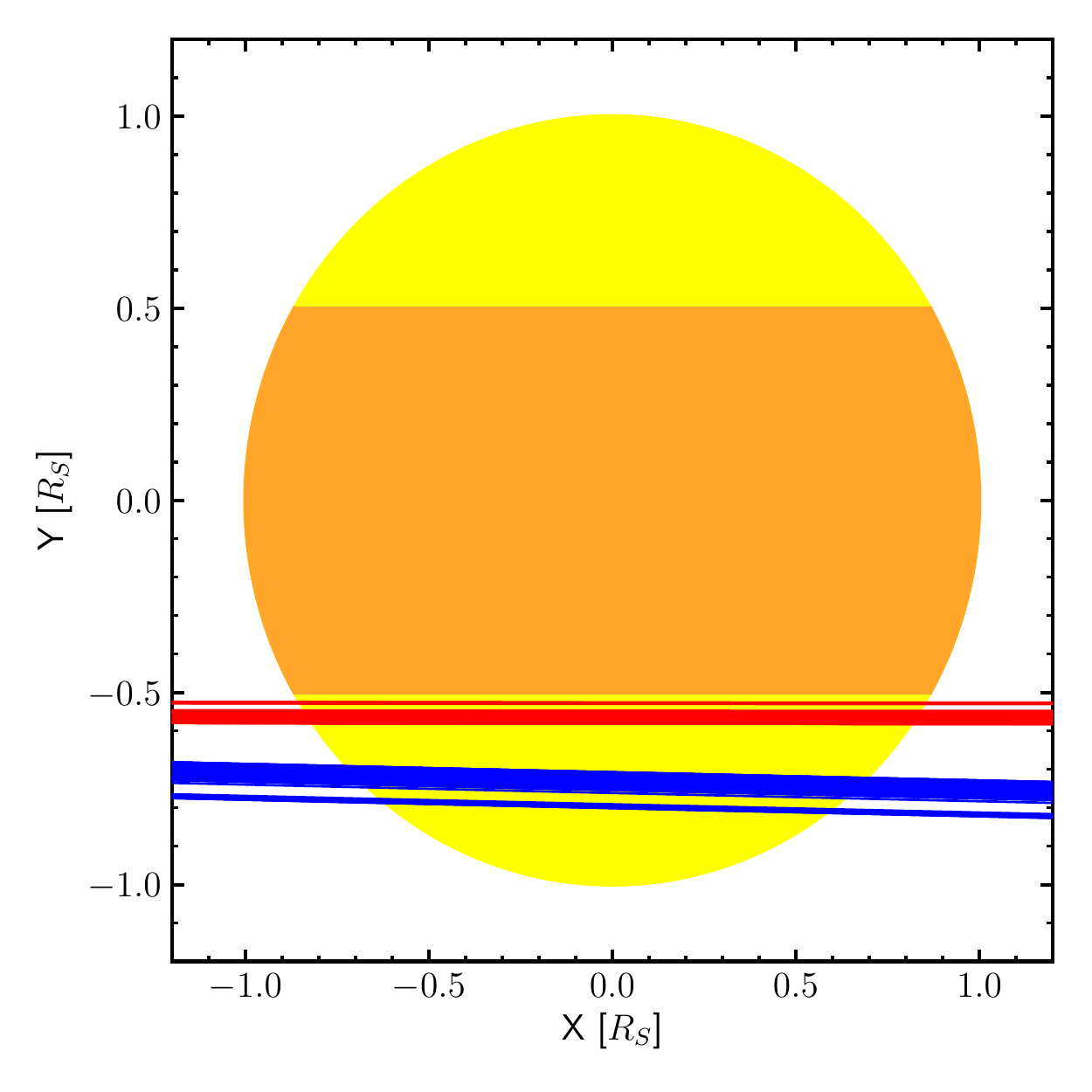}}
\caption{
The latitude coverage of Kepler-9 (yellow circular disc) by all
transit observation of data set III by Kepler-9b (red) and Kepler-9c (blue).
Demonstrated is the best model of data set III. The order of the variations 
can be drawn from Fig.\ref{fig:parameters} or Fig.\ref{fig:parameters_long}
respectively, where the fourth row shows the modelled impact parameters.
The orange area indicates the possible spot occurrence area between $0^\degree$ and $30^\degree$ up- and downwards.}
\label{fig:Latitudes}
\end{figure}

\begin{figure}
\resizebox{\hsize}{!}{\includegraphics{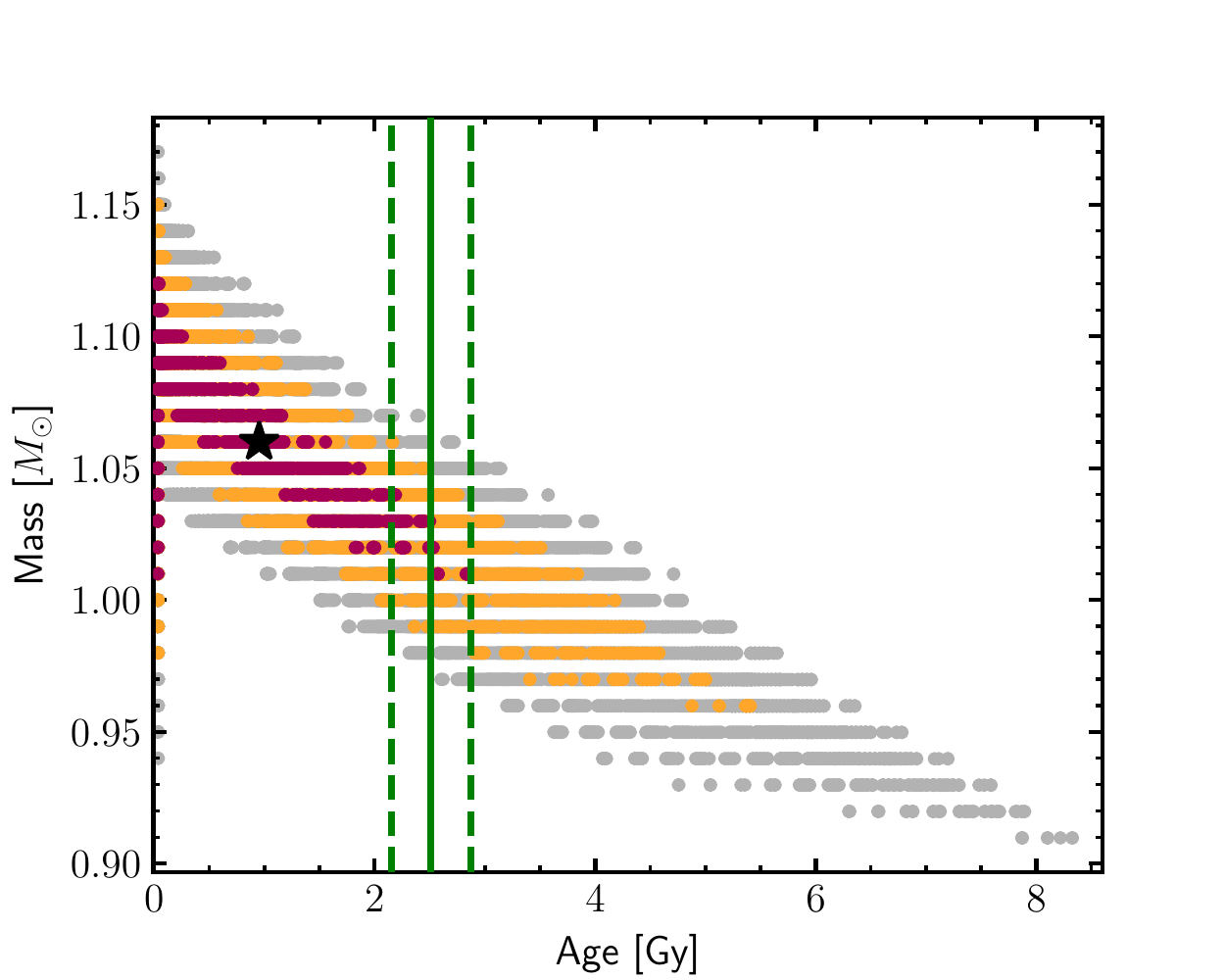}}
\caption{Mass-age diagram of Kepler-9 from MESA stellar evolution
models (MIST). The black star and the red, orange, and grey dots correspond to 
the best matching value and the $1\sigma$, $2\sigma$,
and $3\sigma$ areas derived from results on the density of the data set III
photodynamical modeling and from literature values of the effective temperature,
the surface gravity, and the metallicity by \citet{Holman2010}.
The gyrochronological age is indicated by the green solid line and 
its 1-$\sigma$ range with the green dashed lines.}
\label{fig:MassAge}
\end{figure}
\begin{figure}
\resizebox{\hsize}{!}{\includegraphics{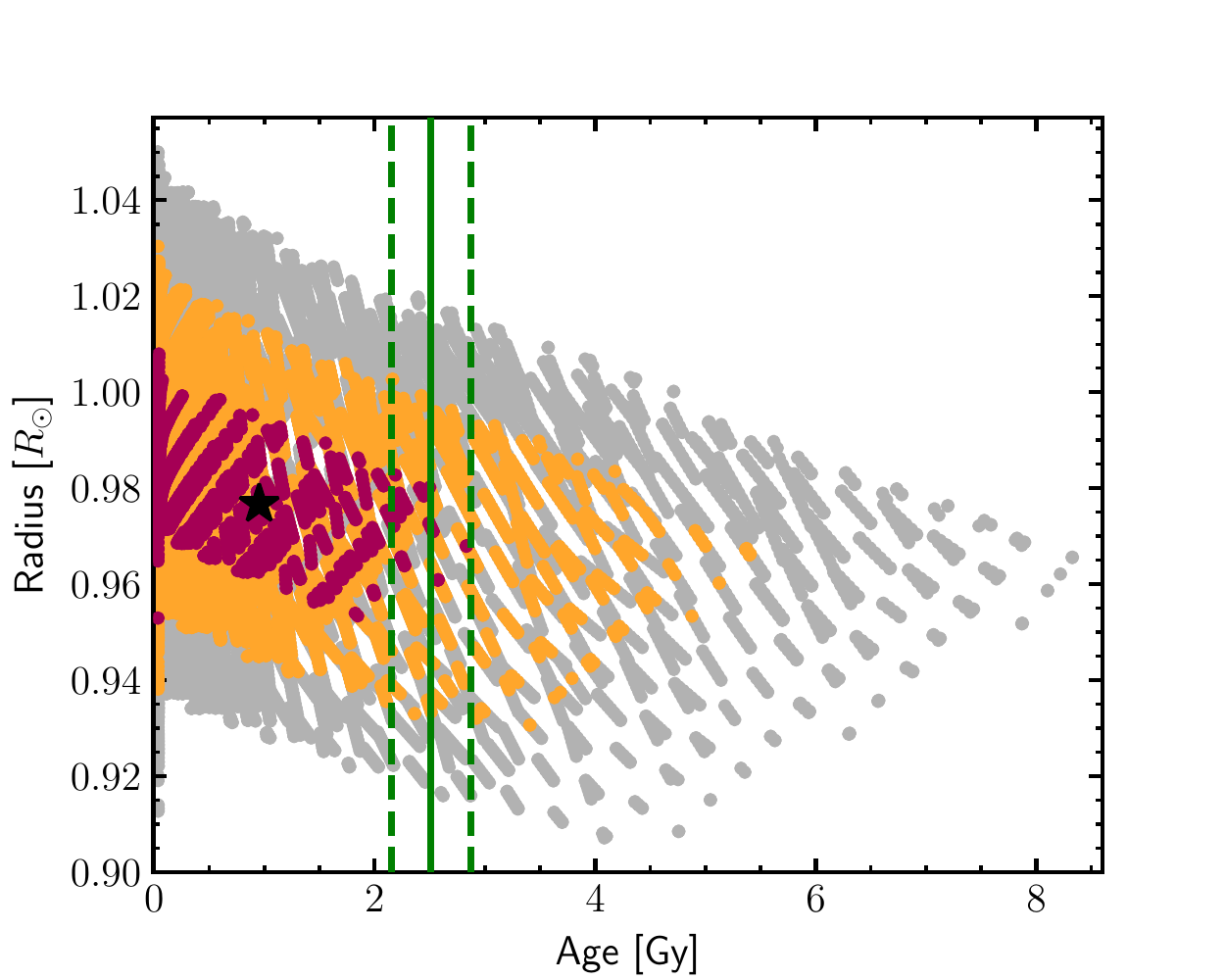}}
\caption{Radius-age diagram of Kepler-9 from MESA stellar evolution
models (MIST). The black star and the red, orange, and grey dots correspond 
to the best matching value and the $1\sigma$, $2\sigma$, and $3\sigma$ areas 
derived from results on the density of the data set III photodynamical 
modeling and from literature values of the effective temperature,
the surface gravity, and the metallicity by \citet{Holman2010}.
The gyrochronological age is indicated by the green solid line and 
its 1-$\sigma$ range with the green dashed lines.}
\label{fig:RadiusAge}
\end{figure}
\begin{figure}
\includegraphics[width=8.5cm]{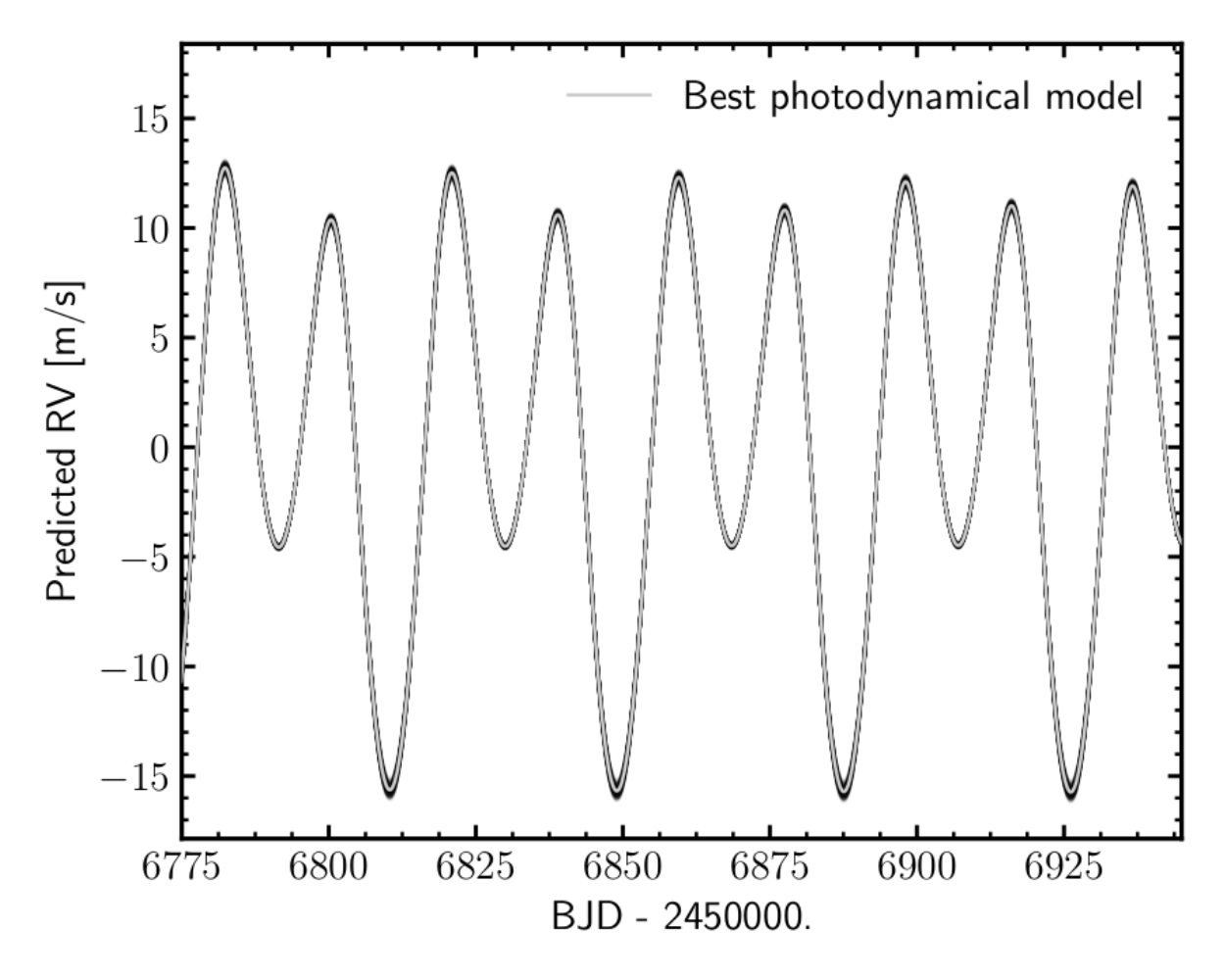}
\caption{Predicted radial velocity measurements from the results 
of the photodynamical modelling of data set III for the approximate 
time span of the new observations listed in the HARPS-N archive}
\label{fig:RV_pred}
\end{figure}

\begin{figure*}
\resizebox{\hsize}{!}{\includegraphics{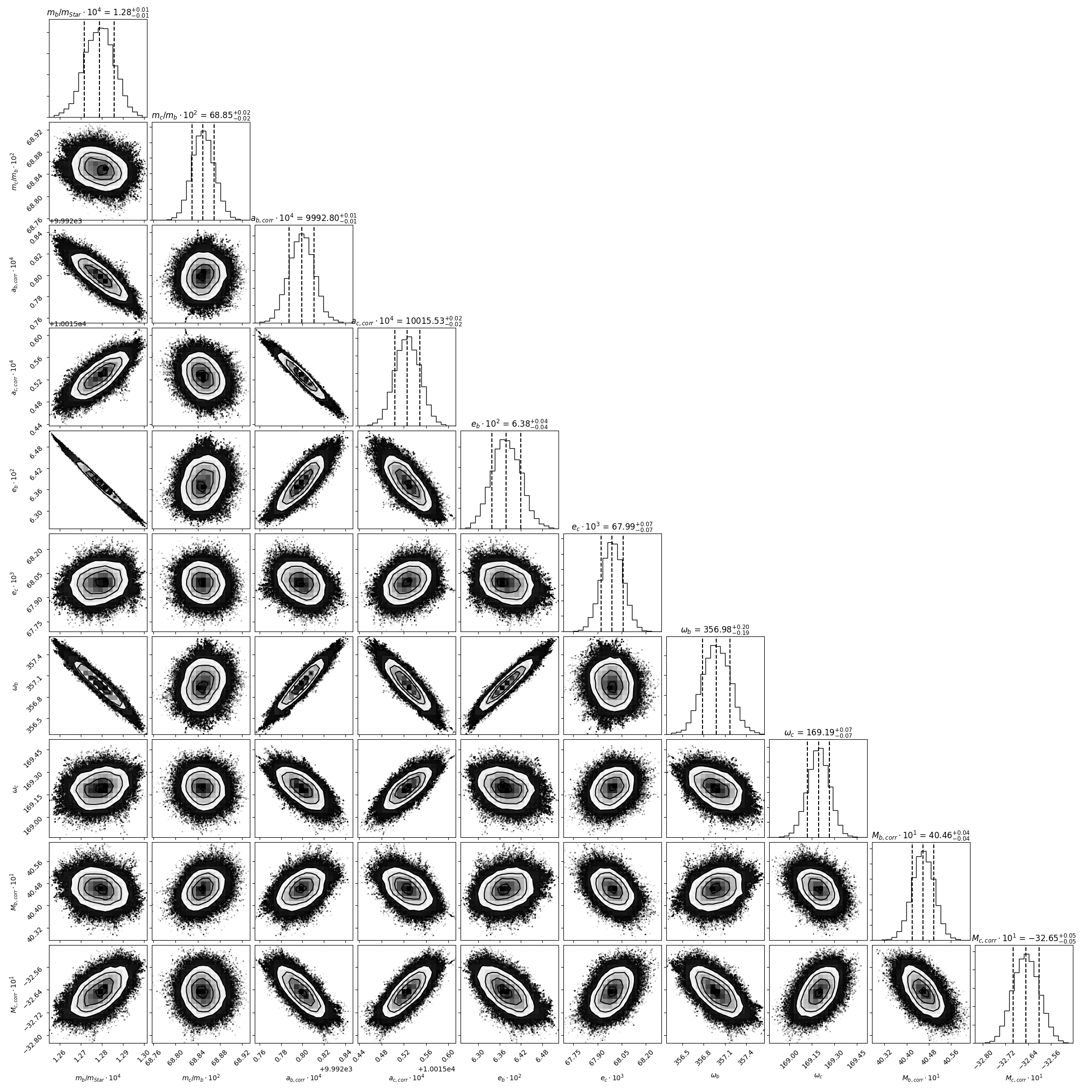}}
\caption{Correlation plot of masses, semi-major-axis, 
eccentricities, longitude of Periastron and mean anomaly from 
MCMC chains of modelling the full dataset. }
\label{fig:corner_mass}
\end{figure*}
\begin{figure*}
\resizebox{\hsize}{!}{\includegraphics{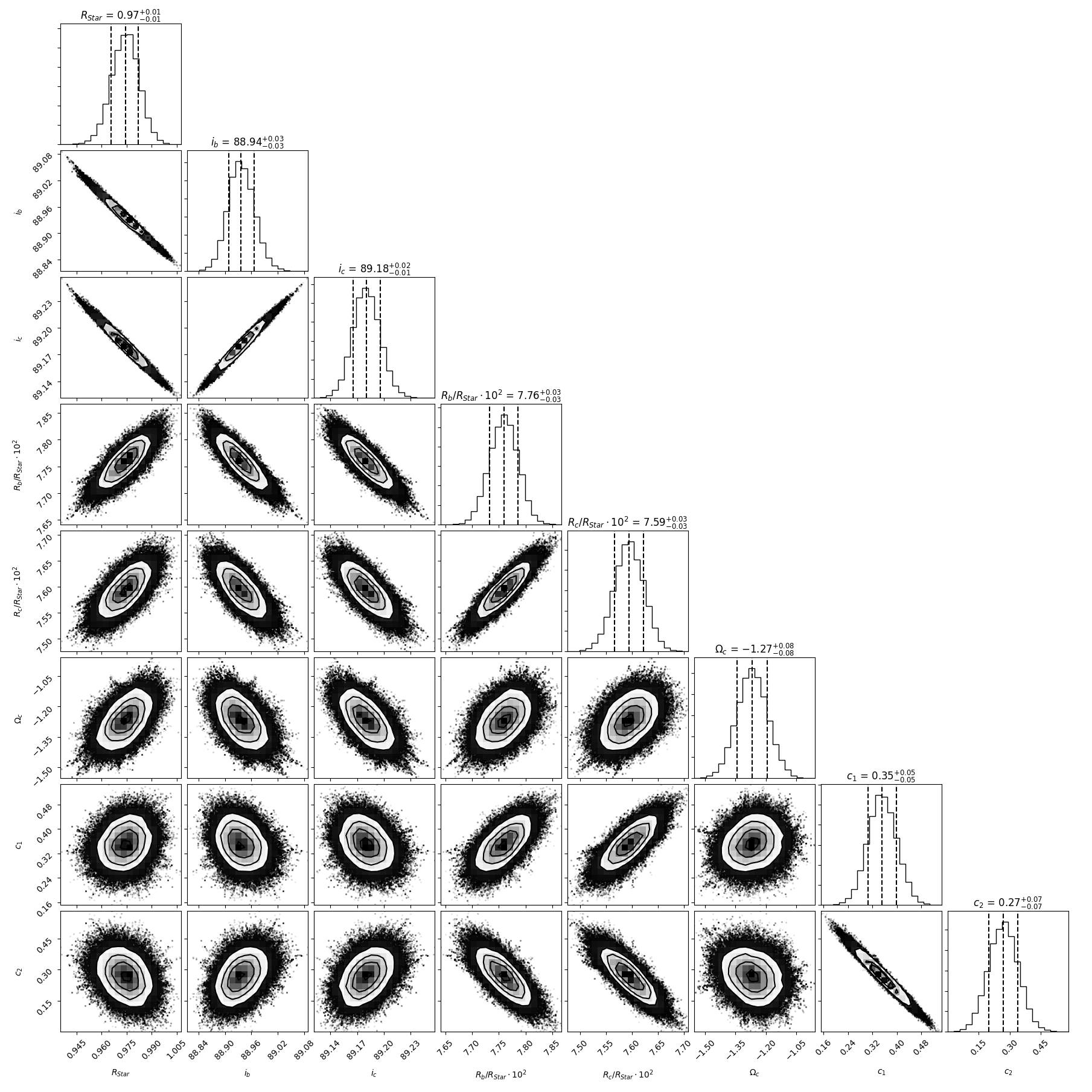}}
\caption{Correlation plot of stellar radius, inclination, 
planetary radii, argument of the ascending node of Kepler-9c 
and limb darkening coefficients from MCMC chains of modelling 
the full dataset. }
\label{fig:corner_radius}
\end{figure*}
\begin{figure*}
\resizebox{\hsize}{!}{\includegraphics{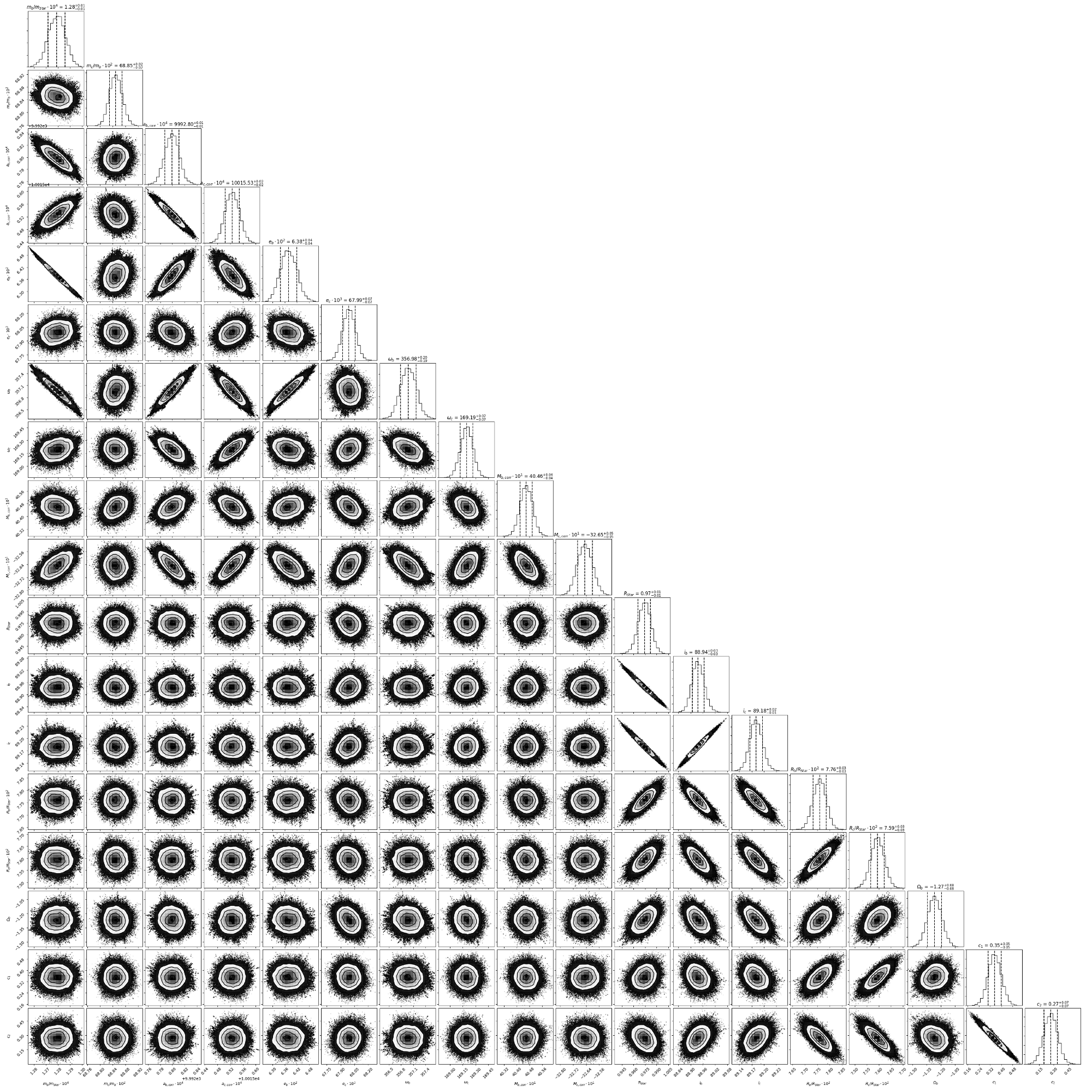}}
\caption{Full correlation plot of all fit parameters from modelling the full dataset.}
\label{fig:corner}
\end{figure*}

\longtab[1]{
\begin{longtable}{cccccccccccc}
\caption{Ephemerides E and transit time predictions in BJD-2400000.0 
from modelling data set III for the next 10 years. The median and 
standard deviation solution of 1000 randomly chosen good models. 
Reference times for ephemeris E=1: $T_b=54977.24962(54)$ and 
$T_c=54969.30566(78)$}\\
\label{table:ttv}
\\
\hline
\hline															
E     &	BJD		 & & E    &	BJD	     & & E    &	BJD		  & & E     &	BJD \\
\hline
\multicolumn{2}{l}{Kepler-9b:} & & &  & & & & & & \\
\hline
$165$ & $58133.0822(7) $ & &$222$ & $59230.2853(6) $ & &$279$ & $60327.6320(13) $ & &$336$ & $61423.9735(30)$ \\
$166$ & $58152.3330(7) $ & &$223$ & $59249.5049(6) $ & &$280$ & $60346.8684(13) $ & &$337$ & $61443.2492(31)$ \\
$167$ & $58171.5797(8) $ & &$224$ & $59268.7261(6) $ & &$281$ & $60366.0990(12) $ & &$338$ & $61462.5362(31)$ \\
$168$ & $58190.8374(9) $ & &$225$ & $59287.9463(7) $ & &$282$ & $60385.3303(12) $ & &$339$ & $61481.8137(31)$ \\
$169$ & $58210.0897(9) $ & &$226$ & $59307.1682(7) $ & &$283$ & $60404.5561(11) $ & &$340$ & $61501.1012(31)$ \\
$170$ & $58229.3543(10) $ & &$227$ & $59326.3890(7) $ & &$284$ & $60423.7831(11) $ & &$341$ & $61520.3792(31)$ \\
$171$ & $58248.6125(10) $ & &$228$ & $59345.6117(7) $ & &$285$ & $60443.0054(11) $ & &$342$ & $61539.6657(31)$ \\
$172$ & $58267.8836(11) $ & &$229$ & $59364.8334(7) $ & &$286$ & $60462.2291(11) $ & &$343$ & $61558.9429(31)$ \\
$173$ & $58287.1474(12) $ & &$230$ & $59384.0569(7) $ & &$287$ & $60481.4489(10) $ & &$344$ & $61578.2269(31)$ \\
$174$ & $58306.4240(12) $ & &$231$ & $59403.2799(7) $ & &$288$ & $60500.6702(10) $ & &$345$ & $61597.5020(30)$ \\
$175$ & $58325.6928(13) $ & &$232$ & $59422.5044(7) $ & &$289$ & $60519.8885(10) $ & &$346$ & $61616.7820(29)$ \\
$176$ & $58344.9740(13) $ & &$233$ & $59441.7290(7) $ & &$290$ & $60539.1082(10) $ & &$347$ & $61636.0537(29)$ \\
$177$ & $58364.2468(13) $ & &$234$ & $59460.9550(7) $ & &$291$ & $60558.3259(10) $ & &$348$ & $61655.3282(28)$ \\
$178$ & $58383.5311(14) $ & &$235$ & $59480.1818(7) $ & &$292$ & $60577.5447(10) $ & &$349$ & $61674.5955(27)$ \\
$179$ & $58402.8069(14) $ & &$236$ & $59499.4096(8) $ & &$293$ & $60596.7623(10) $ & &$350$ & $61693.8635(26)$ \\
$180$ & $58422.0930(14) $ & &$237$ & $59518.6395(8) $ & &$294$ & $60615.9806(10) $ & &$351$ & $61713.1253(25)$ \\
$181$ & $58441.3706(14) $ & &$238$ & $59537.8696(8) $ & &$295$ & $60635.1985(10) $ & &$352$ & $61732.3857(24)$ \\
$182$ & $58460.6569(14) $ & &$239$ & $59557.1034(9) $ & &$296$ & $60654.4168(10) $ & &$353$ & $61751.6414(23)$ \\
$183$ & $58479.9349(14) $ & &$240$ & $59576.3366(9) $ & &$297$ & $60673.6353(10) $ & &$354$ & $61770.8939(22)$ \\
$184$ & $58499.2199(14) $ & &$241$ & $59595.5752(10) $ & &$298$ & $60692.8538(10) $ & &$355$ & $61790.1431(21)$ \\
$185$ & $58518.4971(14) $ & &$242$ & $59614.8123(10) $ & &$299$ & $60712.0730(10) $ & &$356$ & $61809.3878(20)$ \\
$186$ & $58537.7793(14) $ & &$243$ & $59634.0566(11) $ & &$300$ & $60731.2918(10) $ & &$357$ & $61828.6306(20)$ \\
$187$ & $58557.0542(14) $ & &$244$ & $59653.2983(11) $ & &$301$ & $60750.5118(10) $ & &$358$ & $61847.8682(19)$ \\
$188$ & $58576.3322(13) $ & &$245$ & $59672.5491(12) $ & &$302$ & $60769.7310(10) $ & &$359$ & $61867.1050(18)$ \\
$189$ & $58595.6037(13) $ & &$246$ & $59691.7960(13) $ & &$303$ & $60788.9518(10) $ & &$360$ & $61886.3363(17)$ \\
$190$ & $58614.8760(13) $ & &$247$ & $59711.0537(14) $ & &$304$ & $60808.1716(10) $ & &$361$ & $61905.5678(17)$ \\
$191$ & $58634.1429(12) $ & &$248$ & $59730.3063(15) $ & &$305$ & $60827.3932(10) $ & &$362$ & $61924.7941(16)$ \\
$192$ & $58653.4085(12) $ & &$249$ & $59749.5708(16) $ & &$306$ & $60846.6137(10) $ & &$363$ & $61944.0213(16)$ \\
$193$ & $58672.6698(11) $ & &$250$ & $59768.8292(17) $ & &$307$ & $60865.8362(10) $ & &$364$ & $61963.2438(15)$ \\
$194$ & $58691.9279(11) $ & &$251$ & $59788.1003(18) $ & &$308$ & $60885.0576(10) $ & &$365$ & $61982.4676(15)$ \\
$195$ & $58711.1829(10) $ & &$252$ & $59807.3642(18) $ & &$309$ & $60904.2811(10) $ & &$366$ & $62001.6874(15)$ \\
$196$ & $58730.4333(10) $ & &$253$ & $59826.6410(19) $ & &$310$ & $60923.5038(10) $ & &$367$ & $62020.9088(14)$ \\
$197$ & $58749.6817(10) $ & &$254$ & $59845.9098(20) $ & &$311$ & $60942.7283(10) $ & &$368$ & $62040.1269(14)$ \\
$198$ & $58768.9246(9) $ & &$255$ & $59865.1912(21) $ & &$312$ & $60961.9527(10) $ & &$369$ & $62059.3466(14)$ \\
$199$ & $58788.1666(9) $ & &$256$ & $59884.4640(21) $ & &$313$ & $60981.1787(11) $ & &$370$ & $62078.5639(14)$ \\
$200$ & $58807.4026(8) $ & &$257$ & $59903.7488(22) $ & &$314$ & $61000.4054(11) $ & &$371$ & $62097.7826(14)$ \\
$201$ & $58826.6387(8) $ & &$258$ & $59923.0246(22) $ & &$315$ & $61019.6333(11) $ & &$372$ & $62116.9998(14)$ \\
$202$ & $58845.8688(8) $ & &$259$ & $59942.3112(22) $ & &$316$ & $61038.8630(12) $ & &$373$ & $62136.2179(14)$ \\
$203$ & $58865.0999(7) $ & &$260$ & $59961.5887(22) $ & &$317$ & $61058.0934(12) $ & &$374$ & $62155.4354(13)$ \\
$204$ & $58884.3254(7) $ & &$261$ & $59980.8757(23) $ & &$318$ & $61077.3270(13) $ & &$375$ & $62174.6533(13)$ \\
$205$ & $58903.5523(7) $ & &$262$ & $60000.1537(22) $ & &$319$ & $61096.5605(13) $ & &$376$ & $62193.8714(13)$ \\
$206$ & $58922.7744(7) $ & &$263$ & $60019.4395(22) $ & &$320$ & $61115.7989(14) $ & &$377$ & $62213.0895(13)$ \\
$207$ & $58941.9981(7) $ & &$264$ & $60038.7167(22) $ & &$321$ & $61135.0363(15) $ & &$378$ & $62232.3083(13)$ \\
$208$ & $58961.2179(7) $ & &$265$ & $60057.9998(22) $ & &$322$ & $61154.2805(16) $ & &$379$ & $62251.5267(13)$ \\
$209$ & $58980.4393(7) $ & &$266$ & $60077.2749(21) $ & &$323$ & $61173.5225(17) $ & &$380$ & $62270.7463(13)$ \\
$210$ & $58999.6578(7) $ & &$267$ & $60096.5539(21) $ & &$324$ & $61192.7730(18) $ & &$381$ & $62289.9651(13)$ \\
$211$ & $59018.8777(7) $ & &$268$ & $60115.8255(21) $ & &$325$ & $61212.0202(19) $ & &$382$ & $62309.1856(13)$ \\
$212$ & $59038.0957(7) $ & &$269$ & $60135.0989(20) $ & &$326$ & $61231.2777(20) $ & &$383$ & $62328.4050(13)$ \\
$213$ & $59057.3147(7) $ & &$270$ & $60154.3660(19) $ & &$327$ & $61250.5305(21) $ & &$384$ & $62347.6263(13)$ \\
$214$ & $59076.5327(7) $ & &$271$ & $60173.6328(19) $ & &$328$ & $61269.7949(23) $ & &$385$ & $62366.8465(13)$ \\
$215$ & $59095.7514(7) $ & &$272$ & $60192.8943(18) $ & &$329$ & $61289.0534(24) $ & &$386$ & $62386.0687(13)$ \\
$216$ & $59114.9697(7) $ & &$273$ & $60212.1536(17) $ & &$330$ & $61308.3245(25) $ & &$387$ & $62405.2899(13)$ \\
$217$ & $59134.1883(7) $ & &$274$ & $60231.4089(17) $ & &$331$ & $61327.5884(26) $ & &$388$ & $62424.5131(14)$ \\
$218$ & $59153.4073(6) $ & &$275$ & $60250.6603(16) $ & &$332$ & $61346.8653(27) $ & &$389$ & $62443.7355(14)$ \\
$219$ & $59172.6261(6) $ & &$276$ & $60269.9091(15) $ & &$333$ & $61366.1340(28) $ & &$390$ & $62462.9601(14)$ \\
$220$ & $59191.8458(6) $ & &$277$ & $60289.1529(14) $ & &$334$ & $61385.4156(29) $ & &$391$ & $62482.1842(14)$ \\
$221$ & $59211.0649(6) $ & &$278$ & $60308.3953(14) $ & &$335$ & $61404.6884(30) $ & &$392$ & $62501.4102(15)$ \\
\hline
E     &	BJD		 & & E    &	BJD	     & & E    &	BJD		  & & E     &	BJD \\
\hline															
\multicolumn{2}{l}{Kepler-9c:} & & &  & & & & & & \\
\hline
$82$ & $58125.7907(10) $ & & $111$ & $59255.0764(17) $ & & $140$ & $60384.0857(19) $ & & $169$ & $61514.8518(47) $ \\
$83$ & $58164.7408(11) $ & & $112$ & $59294.1486(17) $ & & $141$ & $60423.1271(20) $ & & $170$ & $61553.6402(46) $ \\
$84$ & $58203.6605(12) $ & & $113$ & $59333.2194(17) $ & & $142$ & $60462.1834(21) $ & & $171$ & $61592.4390(44) $ \\
$85$ & $58242.5493(14) $ & & $114$ & $59372.2879(17) $ & & $143$ & $60501.2500(22) $ & & $172$ & $61631.2545(41) $ \\
$86$ & $58281.4089(16) $ & & $115$ & $59411.3529(17) $ & & $144$ & $60540.3229(23) $ & & $173$ & $61670.0929(38) $ \\
$87$ & $58320.2428(18) $ & & $116$ & $59450.4124(17) $ & & $145$ & $60579.3991(24) $ & & $174$ & $61708.9592(34) $ \\
$88$ & $58359.0557(20) $ & & $117$ & $59489.4635(17) $ & & $146$ & $60618.4764(25) $ & & $175$ & $61747.8572(30) $ \\
$89$ & $58397.8536(21) $ & & $118$ & $59528.5028(16) $ & & $147$ & $60657.5537(25) $ & & $176$ & $61786.7885(27) $ \\
$90$ & $58436.6427(22) $ & & $119$ & $59567.5263(16) $ & & $148$ & $60696.6301(25) $ & & $177$ & $61825.7525(25) $ \\
$91$ & $58475.4297(22) $ & & $120$ & $59606.5294(15) $ & & $149$ & $60735.7055(25) $ & & $178$ & $61864.7467(24) $ \\
$92$ & $58514.2212(21) $ & & $121$ & $59645.5080(16) $ & & $150$ & $60774.7795(25) $ & & $179$ & $61903.7666(25) $ \\
$93$ & $58553.0240(20) $ & & $122$ & $59684.4584(17) $ & & $151$ & $60813.8522(25) $ & & $180$ & $61942.8073(26) $ \\
$94$ & $58591.8444(18) $ & & $123$ & $59723.3785(20) $ & & $152$ & $60852.9232(24) $ & & $181$ & $61981.8635(28) $ \\
$95$ & $58630.6884(17) $ & & $124$ & $59762.2678(23) $ & & $153$ & $60891.9918(24) $ & & $182$ & $62020.9307(29) $ \\
$96$ & $58669.5608(15) $ & & $125$ & $59801.1279(26) $ & & $154$ & $60931.0567(24) $ & & $183$ & $62060.0046(30) $ \\
$97$ & $58708.4650(13) $ & & $126$ & $59839.9620(29) $ & & $155$ & $60970.1159(23) $ & & $184$ & $62099.0820(31) $ \\
$98$ & $58747.4020(12) $ & & $127$ & $59878.7749(32) $ & & $156$ & $61009.1667(23) $ & & $185$ & $62138.1607(32) $ \\
$99$ & $58786.3710(12) $ & & $128$ & $59917.5724(33) $ & & $157$ & $61048.2058(22) $ & & $186$ & $62177.2393(32) $ \\
$100$ & $58825.3689(12) $ & & $129$ & $59956.3608(34) $ & & $158$ & $61087.2292(21) $ & & $187$ & $62216.3170(32) $ \\
$101$ & $58864.3913(13) $ & & $130$ & $59995.1465(34) $ & & $159$ & $61126.2324(21) $ & & $188$ & $62255.3933(32) $ \\
$102$ & $58903.4329(14) $ & & $131$ & $60033.9364(33) $ & & $160$ & $61165.2112(22) $ & & $189$ & $62294.4682(32) $ \\
$103$ & $58942.4889(15) $ & & $132$ & $60072.7371(32) $ & & $161$ & $61204.1622(24) $ & & $190$ & $62333.5415(32) $ \\
$104$ & $58981.5547(16) $ & & $133$ & $60111.5551(29) $ & & $162$ & $61243.0830(28) $ & & $191$ & $62372.6128(32) $ \\
$105$ & $59020.6264(16) $ & & $134$ & $60150.3963(27) $ & & $163$ & $61281.9731(32) $ & & $192$ & $62411.6815(31) $ \\
$106$ & $59059.7012(17) $ & & $135$ & $60189.2658(24) $ & & $164$ & $61320.8338(37) $ & & $193$ & $62450.7463(30) $ \\
$107$ & $59098.7772(17) $ & & $136$ & $60228.1670(21) $ & & $165$ & $61359.6686(41) $ & & $194$ & $62489.8054(30) $ \\
$108$ & $59137.8532(17) $ & & $137$ & $60267.1014(19) $ & & $166$ & $61398.4818(44) $ & & &\\
$109$ & $59176.9285(17) $ & & $138$ & $60306.0681(18) $ & & $167$ & $61437.2792(46) $ & &&\\
$110$ & $59216.0030(17) $ & & $139$ & $60345.0643(18) $ & & $168$ & $61476.0670(47) $ & &&\\
\hline															
\end{longtable}
}

\end{appendix}

\end{document}